\documentclass[twocolumn]{aastex63}

\newcommand{\NstarsTooFar}{52}
\newcommand{\Nstarsobs}{1157}
\newcommand{\Nsystems}{1083}
\newcommand{\NstarsGaia}{1005}
\newcommand{\NstarsSIMBAD}{78}
\newcommand{\NstarsSIMBADSpT}{1038}
\newcommand{\NstarsGaiaSpT}{45}

\newcommand{\Ncomps}{301}
\newcommand{\NcompsRAOprims}{74}
\newcommand{\NcompsRAO}{154}
\newcommand{\NcompsGaiawide}{73}
\newcommand{\NcompsGaiaDet}{147}

\newcommand{\Ncompsbound}{200}
\newcommand{\NcompsboundRAO}{53}
\newcommand{\Ncompsbckgrd}{2}
\newcommand{\Ncompsunknown}{99}
\newcommand{\Nbinaries}{255}

\newcommand{\Ntriples}{19}
\newcommand{\Nquads}{2}
\newcommand{\NhigherOrdersTight}{4}
\newcommand{\NhigherOrdersWide}{4}
\newcommand{\NhigherOrdersBoth}{13}
\newcommand{\NcompsElBadry}{158}
\newcommand{\NSBs}{70}
\newcommand{\NEBs}{4}
\newcommand{\NSBsnewBinaries}{42}
\newcommand{\NSBsnewHOs}{28}
\newcommand{\NcompsUNRESOLVED}{94}
\newcommand{\NhigherOrdersUNRESOLVED}{27}
\newcommand{\NcompsNew}{49}

\newcommand{\NstarsAcc}{727}
\newcommand{\NstarsAccSig}{191}
\newcommand{\NstarsAccNOsig}{536}
\newcommand{\NstarsAccNOComps}{540}
\newcommand{\NstarsAccNOCompsSig}{76}
\newcommand{\NstarsAccNOCompsNOsig}{464}

\newcommand{\NstarsAccCompsSig}{115}
\newcommand{\NstarsAccCompsNOsig}{72}
\newcommand{\NstarsAccWideComps}{112}
\newcommand{\NstarsAccWideCompsSig}{47}
\newcommand{\NstarsAccWideCompsNOsig}{65}
\newcommand{\NstarsAccTightComps}{75}
\newcommand{\NstarsAccTightCompsSig}{68}
\newcommand{\NstarsAccTightCompsNOsig}{7}

\newcommand{\FracAccNOCompsSig}{14.1\%$\pm$1.6\%}
\newcommand{\FracAccNOCompsNOsig}{85.9\%$\pm$4.0\%}
\newcommand{\FracAccWideCompsSig}{42.0\%$\pm$6.1\%}
\newcommand{\FracAccWideCompsNOsig}{58.0\%$\pm$7.2\%}
\newcommand{\FracAccTightCompsSig}{91$_{-4}^{+2}$\%}
\newcommand{\FracAccTightCompsNOSig}{9$_{-2}^{+5}$\%}

\newcommand{\MultFracALL}{25.5\%$\pm$1.5\%}
\newcommand{\MultFracHOAll}{1.9\%$\pm$0.4\%}

\newcommand{\MultFracA}{25$_{-10}^{+27}$\%}
\newcommand{\MultFracF}{40$_{-5}^{+6}$\%}
\newcommand{\MultFracG}{26.9\%$\pm$4.8\%}
\newcommand{\MultFracK}{27.8\%$\pm$3.2\%}
\newcommand{\MultFracM}{20.8\%$\pm$2.0\%}
\newcommand{\MultFracWD}{7$_{-2}^{+8}$\%}
\newcommand{\MultFracFGK}{29.5\%$\pm$2.5\%}
\newcommand{\MultFracMunrs}{28.2\%$\pm$2.3\%}
\newcommand{\MultFracFGKunrs}{40.9\%$\pm$3.0\%}
\newcommand{\MultFracMSBs}{21.7\%$\pm$2.1\%}
\newcommand{\MultFracFGKSBs}{37.1\%$\pm$2.9\%}

\newcommand{\MultFracHOA}{25$_{-10}^{+27}$\%}

\newcommand{\MultFracHOG}{2.5\%$\pm$1.5\%}
\newcommand{\MultFracHOK}{0.8\%$\pm$0.5\%}
\newcommand{\MultFracHOM}{2.5\%$\pm$0.7\%}

\newcommand{\MultFracHOFGK}{1.1\%$\pm$0.5\%}
\newcommand{\MultFracHOMunrs}{3.9\%$\pm$0.9\%}
\newcommand{\MultFracHOFGKunrs}{5.5\%$\pm$1.1\%}
\newcommand{\MultFracHOMSBs}{3.5\%$\pm$0.8\%}
\newcommand{\MultFracHOFGKSBs}{6.2\%$\pm$1.2\%}

\newcommand{\NstarsRECOVrij}{77}
\newcommand{\NcompsRECOVrij}{15}
\newcommand{\NcompsRECOVrijunr}{5}
\newcommand{\NnocompsRECOVrijunr}{9}
\newcommand{\NcompsRag}{51}
\newcommand{\NcompsWinters}{74}
\newcommand{\NcompsWDS}{110}
\newcommand{\NcompsIntFourth}{17}
\newcommand{\NcompsCats}{252}

\usepackage{txfonts}
\usepackage{graphicx}
\usepackage{multirow}
\usepackage{rotating}
\usepackage{verbatim}
\usepackage{url}
\usepackage{dcolumn}
\usepackage[utf8]{inputenc}

\graphicspath{{./}{figures/}}
\usepackage{float}
\usepackage[caption = false]{subfig}
\usepackage{hyperref}

\newcommand\new[1]{{\color{blue}#1}}          
\renewcommand\new[1]{\color{black}#1}

\shorttitle{Robo25 Survey}
\shortauthors{Salama et al.}

\begin{document}

\title{An Adaptive Optics Census of Companions to Northern Stars Within 25 pc with Robo-AO}

\correspondingauthor{Ma\"issa Salama}
\email{msalama@ucsc.edu}

\author[0000-0002-5082-6332]{Ma\"issa Salama}
\altaffiliation{Now at Astronomy \& Astrophysics Department, University of California, Santa Cruz, CA 95064, USA}
\affiliation{Institute for Astronomy, University of Hawai`i at M\={a}noa, Hilo, HI 96720, USA}

\author[0000-0002-0619-7639]{Carl Ziegler}
\affiliation{Department of Physics, Engineering and Astronomy, Stephen F. Austin State University, 1936 North St, Nacogdoches, TX 75962, USA}

\author[0000-0002-1917-9157]{Christoph Baranec}
\affiliation{Institute for Astronomy, University of Hawai`i at M\={a}noa, Hilo, HI 96720, USA}

\author[0000-0003-2232-7664]{Michael C. Liu}
\affiliation{Institute for Astronomy, University of Hawai`i at M\={a}noa, Honolulu, HI 96822, USA}

\author[0000-0001-9380-6457]{Nicholas M. Law}
\affiliation{Department of Physics and Astronomy, University of North Carolina at Chapel Hill, Chapel Hill, NC 27599-3255, USA}

\author[0000-0002-0387-370X]{Reed Riddle}
\affiliation{Division of Physics, Mathematics, and Astronomy, California Institute of Technology, Pasadena, CA 91125, USA}

\author[0000-0002-9061-2865]{Todd J. Henry}
\affiliation{RECONS Institute, Chambersburg, PA 17201, USA}

\author[0000-0001-6031-9513]{Jennifer G. Winters}
\affil{Center for Astrophysics $\vert$ Harvard \& Smithsonian, 60 Garden Street, Cambridge, MA 02138, USA}

\author[0000-0003-0193-2187]{Wei-Chun Jao}
\affiliation{Department of Physics and Astronomy, Georgia State University, Atlanta, GA 30302-4106, USA}

\author[0000-0002-8439-7767]{James Ou}
\affiliation{Institute for Astronomy, University of Hawai`i at M\={a}noa, Hilo, HI 96720, USA}

\author[0000-0002-3376-7297]{Arcelia Hermosillo Ruiz}
\affiliation{Astronomy \& Astrophysics Department, University of California, Santa Cruz, CA 95064, USA}

\begin{abstract}

In order to assess the multiplicity statistics of stars across spectral types and populations in a volume-limited sample, we censused nearby stars for companions with Robo-AO. We report on observations of \Nstarsobs\ stars of all spectral types within 25~pc with decl. $>-13^{\circ}$ searching for tight companions. We detected \NcompsRAO\ companion candidates with separations ranging from $\sim$0$\farcs$15 to 4$\farcs$0 and magnitude differences up to $\Delta$m$_{\textit{i'}}\le$7 using the robotic adaptive optics instrument Robo-AO. We confirmed physical association from Gaia EDR3 astrometry for \NcompsboundRAO\ of the companion candidates, \Ncompsunknown\ remain to be confirmed, and \Ncompsbckgrd\ were ruled out as background objects. We complemented the high-resolution imaging companion search with a search for co-moving objects with separations out to 10,000 AU in Gaia EDR3, which resulted in an additional \NcompsGaiaDet\ companions \new{registered}. Of the \Ncomps\ total companions reported in this study, \NcompsNew\ of them are new discoveries. Out of the \NstarsAccSig\ stars with significant acceleration measurements in the Hipparcos--Gaia catalog of accelerations, we detect companions around \NstarsAccCompsSig\ of them, with the significance of the acceleration increasing as the companion separation decreases. From this survey, we report the following multiplicity fractions (compared to literature values): \MultFracFGKunrs\ (44\%) for FGK stars and \MultFracMunrs\ (27\%) for M stars, as well as higher-order fractions of \MultFracHOFGKunrs\ (11\%) and \MultFracHOMunrs\ (5\%) for FGK stars and M type stars, respectively. 

\end{abstract}


\section{Introduction}
\label{sec:intro}

Stellar multiples provide important constraints for stellar formation and evolution models through detailed statistics of their multiplicity rates as well as detailed studies of individual systems. Studying the orbits of multiple star systems may yield information on the intrinsic properties of the constituent stars (e.g., mass ratios, separation distributions), improving our knowledge of fundamental astrophysical links such as the mass-luminosity relationship, dynamical interactions, and molecular cloud formation environments (see \citealt{Bate15} review paper). Studying companions around the nearest stars is also vital to the current and next generation of transiting-planet hunting missions, allowing quick identification of false-positive transit signals and calibrating estimates of planetary characteristics (e.g., \citealt{Ziegler18}).

The stars in our solar neighborhood are among the most studied in astronomy. Their proximity allows even very faint stars to be counted, providing the best census of our galaxy's stellar population. Various studies have been conducted, focusing on different populations of nearby stars. For low-mass stars, \cite{Ward-Duong15} surveyed 245 late-K to mid-M stars within 15~pc combining adaptive optics imaging and digitized wide-field archival plates and reported 65 companions. \cite{Winters19} performed a volume-limited stellar multiplicity study of 1120 M dwarfs within 25~pc, combining observations with a comprehensive literature search and report on the multiplicity rates and companion separation distribution trends. \cite{Lamman20} observed 5566 nearby M type stars with adaptive optics imaging and detect 553 tight stellar companion candidates. \cite{Vrijmoet20} report on 542 M type stars within 25~pc, comparing 20 years of astrometric data with Gaia DR2 and define criteria for finding unresolved companions in Gaia DR2 data. On the very low-mass end of stars, \cite{Gagliuffi19} conducted a volume-limited spectroscopic study of ultracool dwarfs (M7--L5) within 25~pc and report on spectral binary statistics.

Focusing on higher-mass FGK stars, \cite{Raghavan10} presents the results from a sample of 454 stars within 25~pc. They report detailed statistics on stellar multiplicity fractions and trends with stellar properties (such as mass, age, metallicity). \cite{Tokovinin14} and \cite{Riddle15} focused on the higher-order multiplicity of F and G stars within 67~pc. Others such as \cite{Kervella19} have used Gaia to search for companions within 50~pc using changes in proper motion between the Hipparcos and Gaia catalogs and \cite{Rebassa21} searched for systems consisting of white-dwarf and main sequence pairs within 100~pc.

For the Robo-AO Solar Neighborhood Survey, we observed a volume-limited sample of $>$1200 stars across all spectral types within 25 pc, made possible by the observing efficiency provided by the fully-autonomous laser adaptive optics (AO) Robo-AO instrument. 
Our goal was to detect otherwise unresolved companions in order to estimate the multiplicity fraction of stars across spectral types and provide a census of companions inaccessible to non-AO observations.

In Section \ref{sec:targetselection} we describe the survey target selection and observations, followed by the data reduction in Section \ref{sec:datareduction}. In Section \ref{sec:results} we report the companion detections and characterizations, and a complementary search for wide companions with Gaia. This is followed by a discussion of Gaia companion recovery, accelerating stars, higher-order multiple systems, and multiplicity fractions by spectral types in Section \ref{sec:discussion}, and we conclude in Section \ref{sec:conclusion}.

\section{Survey Targets and Observations}
\label{sec:targetselection}

\subsection{Target Selection}
A volume-limited survey requires a well vetted target list. Our list is based on targets initially identified by the REsearch Consortium On Nearby Stars (RECONS, \url{www.recons.org}) collaboration, whose goal is to discover and characterize all stars and their environments within 25~pc. To consider stars as nearby stars within 25 pc, prior to the Gaia data releases, RECONS calculated the weighted mean trigonometric parallaxes from the literature or published astrometry catalogs, i.e., Hipparcos catalog \citep{vanLeeuwen07} or Yale Parallax Catalog \citep{vanAltena95}. For stars without trigonometric parallaxes, their distances were estimated using photographic plate photometry, and optical and/or near-IR photometry (see \citealt{Winters15}, \citealt{Boyd11} and references therein). Our targets have declinations $>-13^{\circ}$ and $0 < V < 16$ magnitude range.

\subsection{Observations}
We obtained high-angular-resolution images of 1221 stars during 58 nights of observations between 2012 June 17 and 2013 October 23. The observations were performed using the automated Robo-AO laser adaptive optics system \citep{baranec13, baranec14, riddle12} mounted on 1.5-m telescope at Palomar observatory. The adaptive optics system operates at a control loop rate of 1.2 kHz to correct high-order wavefront aberrations. Image displacement is corrected post-facto using the target star in the science images that are captured at a rate of 8.6 frames per second. Observations were taken in the \textit{i}\textsuperscript{$\prime$}-band filter. 

Typical seeing at the Palomar Observatory is between 0$\farcs$8 and 1$\farcs$8, with median around 1$\farcs$1 \citep{Cenko06}. The typical FWHM (diffraction limited) resolution of the Robo-AO system is 0$\farcs$15. Specifications of the Robo-AO solar neighborhood survey are summarized in Table$~\ref{tab:specs}$.

\begin{table}
\renewcommand{\arraystretch}{1.3}
\begin{longtable}{ll}
\caption{\label{tab:specs}The specifications of the Robo-AO solar neighborhood survey}
\\
\hline
Field size & 44\arcsec $\times$ 44\arcsec\\
Detector format       	& 1024$^2$ pixels\\
Pixel scale & 43.1 mas / pix\\
Exposure time per target & 120 seconds \\
Targets observed / hour & 20\\
Observation dates & 2012 Jun 17 -- 2013 Oct 23\\
\hline
\label{tab:specs}
\end{longtable}
\end{table}

\subsection{Target Verification}
\label{sec:targetverification}
To verify that the star viewed in the image is the desired target, we created Digital Sky Survey and PanSTARRS cutouts of similar angular size around the target coordinates. Each image was manually inspected to confirm the target observed with Robo-AO was the correct star, based on surrounding stars and/or brightness, as well as assess image quality. Images of stars were rejected if the wrong star was observed or if there had been a clear issue during the observation, i.e. telescope moved or AO correction turned off during integration. Approximately 98$\%$ of our images passed inspection, and for all but five of the rejected images, additional images of the same target were available. 

All verified targets were cross-matched with Gaia EDR3 to obtain their parallaxes. Three stars were found to have distances between 50 and 500 pc and we are unable to confirm the distances for another four stars. These observations are thus omitted from this paper. Fifty two stars (4.3\% of our sample) were initially thought to be within 25~pc, but are now also known to be more distant according to the Gaia parallax measurements. Of those \NstarsTooFar, 87\% (45 stars) are within 30~pc and 7 stars extend out to distances of 30--40~pc. These stars and their 19 companions (8 detected within 4$\arcsec$ with Robo-AO and 11 found in the Gaia wide co-moving search, see \S\ref{sec:GaiaComove}) were removed from the analysis throughout this paper as they are beyond our distance cut-off of 25~pc but the ones with companions are listed in the Appendix Table \ref{tab:beyond25pc}.

After removing stars with distances greater than 25pc, we are left with \Nstarsobs\ stars confirmed to be within 25~pc. Of these, \NcompsRAOprims\ are wide binaries where we observed each component of the binary separately with Robo-AO. So we observed \Nstarsobs\ stars, which make up \Nsystems\ unique systems.

\begin{figure}
\centering
\includegraphics[width=240pt]{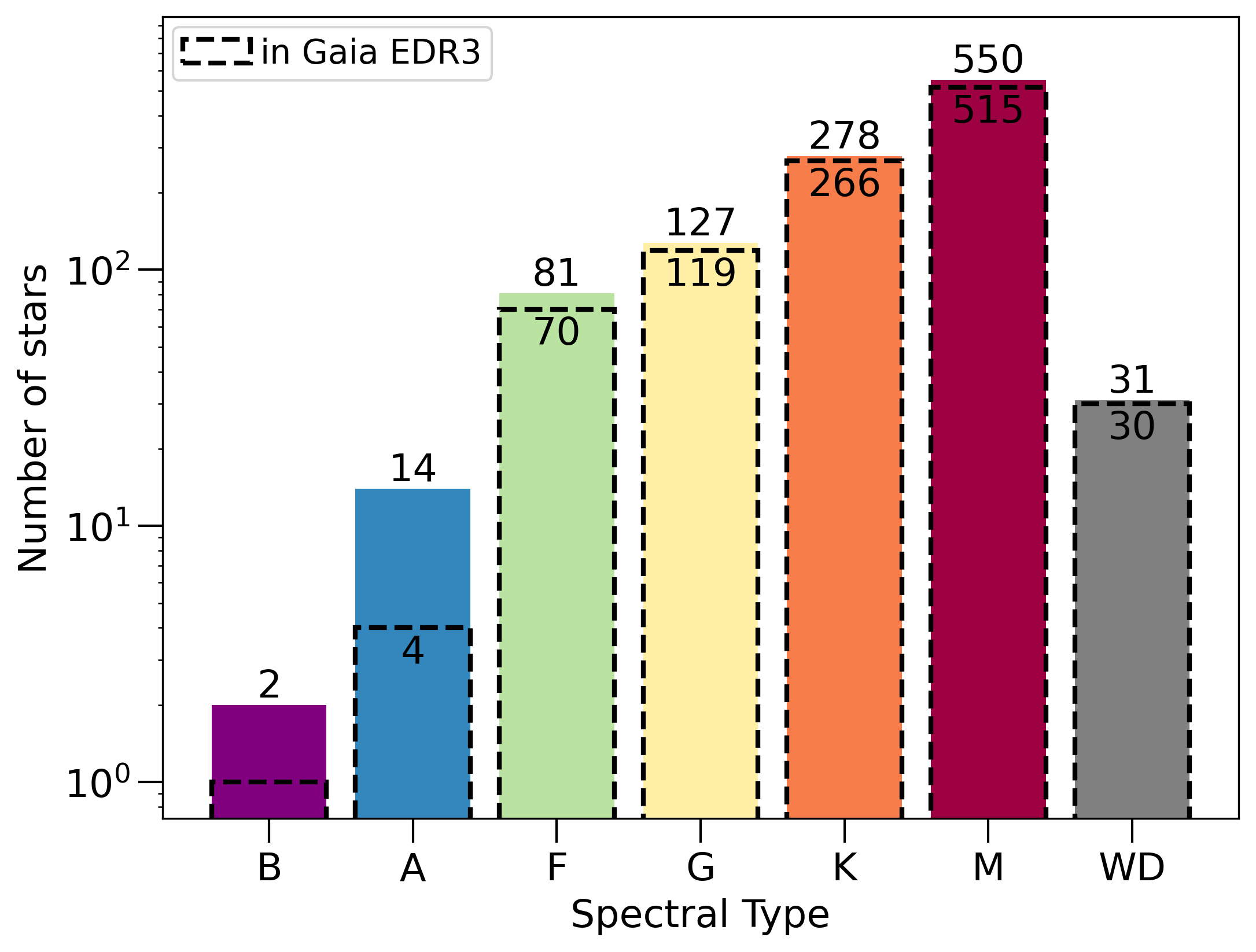}
\caption{Spectral type distribution of observed stars making up the Robo-AO Solar Neighborhood Survey sample. The dashed lines and corresponding numbers are the stars in our sample with Gaia EDR3 parallaxes. Almost half of our sample are M type stars.}
\label{fig:targets}
\end{figure}

Therefore, the sample of stars reported throughout this paper consists of \Nsystems\ primary stars (summarized in Table \ref{tab:targets}), of which \NstarsGaia\ have Gaia EDR3 information. The distances for the remaining \NstarsSIMBAD\ stars were obtained from SIMBAD\new{, all but one (RAO~0739+0513) with parallax measurements}. All observations and distance references are listed in Table~\ref{tab:obs_all} in the Appendix. The spectral type estimates are from SIMBAD when available (\NstarsSIMBADSpT\ of the stars) and the remaining \NstarsGaiaSpT\ not found in SIMBAD are derived from Gaia EDR3 colors (Bp--Rp and G--Rp) and absolute G magnitudes. Using the \cite{Pecaut13} color-Teff grids\footnote{\url{http://www.pas.rochester.edu/~emamajek/EEM_dwarf_UBVIJHK_colors_Teff.txt} (version 2021.03.02)} we estimated the spectral types for each color and magnitude and averaged them to get our final spectral type estimate. For systems with sub-arcsecond companions, the spectral type is a composite type that includes both components of the binary. The spectral type distribution of stars observed and reported in this paper are shown in Figure~\ref{fig:targets}, denoting the sample with Gaia EDR3 information. Many of the early-type stars are not in the Gaia sample because they are too bright (with $V$ magnitudes spanning 0 to 3.5) and therefore do not have Gaia parallax measurements. Furthermore, thirteen stars are identified as young ($\lesssim$ 300 Myrs) low-mass (K or M type) stars by the LASSO survey \citep{Salama21}.

\begin{table}[h]
\centering
\caption{Summary of Observed Stars
    \label{tab:obs_lists_summary}}
\begin{tabular}{lc}
\hline
\textbf{Total observed:} & 1221 stars \\
\hline
\quad rejected during target verification & 5 \\
\quad unable to confirm distance & 4 \\
\quad $>$ 50 pc & 3 \\
\quad 30 -- 40 pc & 7 \\
\quad 25 -- 30 pc & 45 \\
\hline
\textbf{Final sample $<$ 25 pc:} & 1157 individual stars \\
& 1083 systems \\
\hline
\label{tab:targets}
\end{tabular}
\end{table}

\subsection{Volume-Limited Sample Completeness}
\label{sec:completeness}

\begin{figure}
\centering
\subfloat[]{\includegraphics[width=245pt]{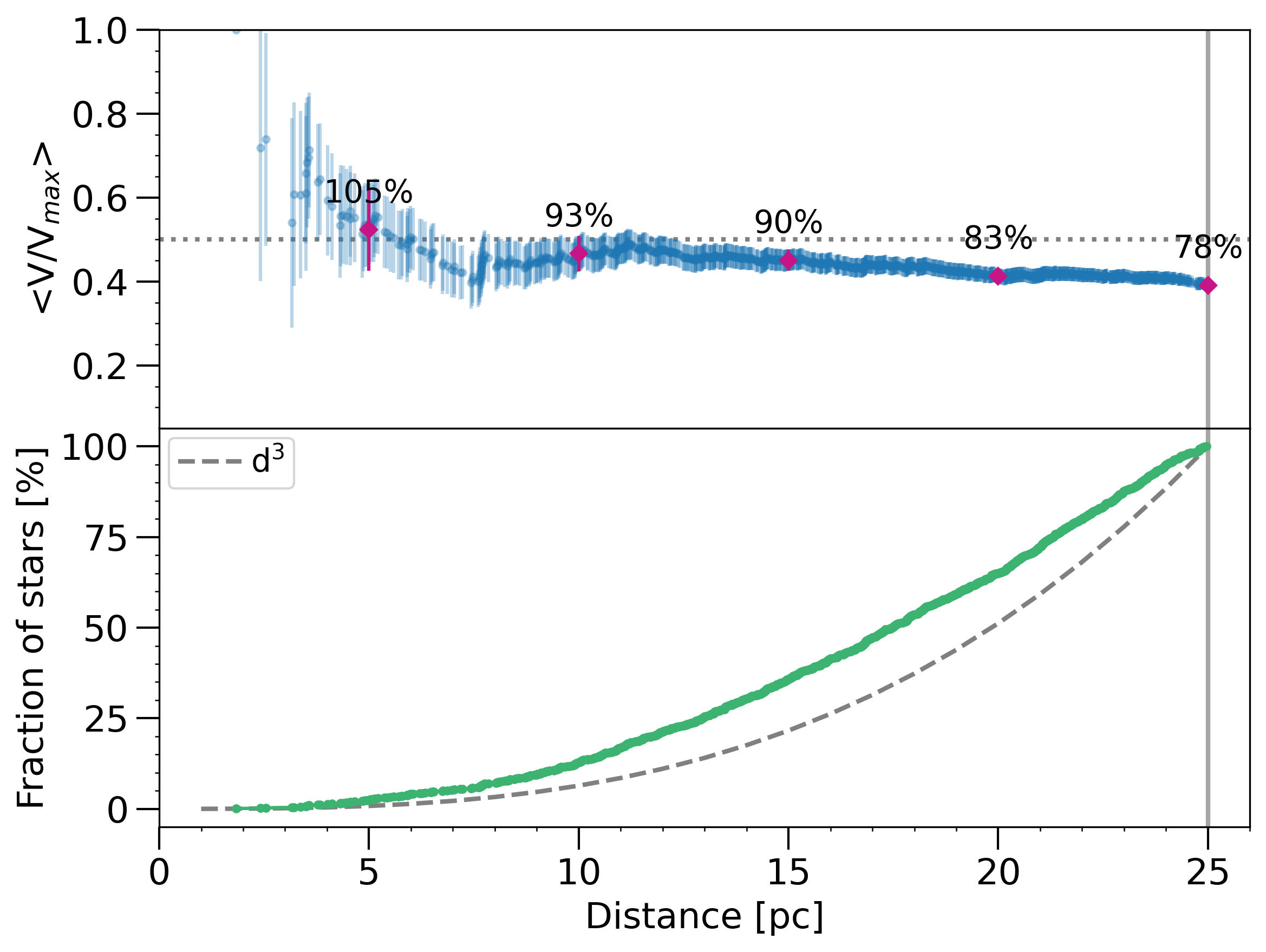}}\\
\subfloat[]{\includegraphics[width=235pt]{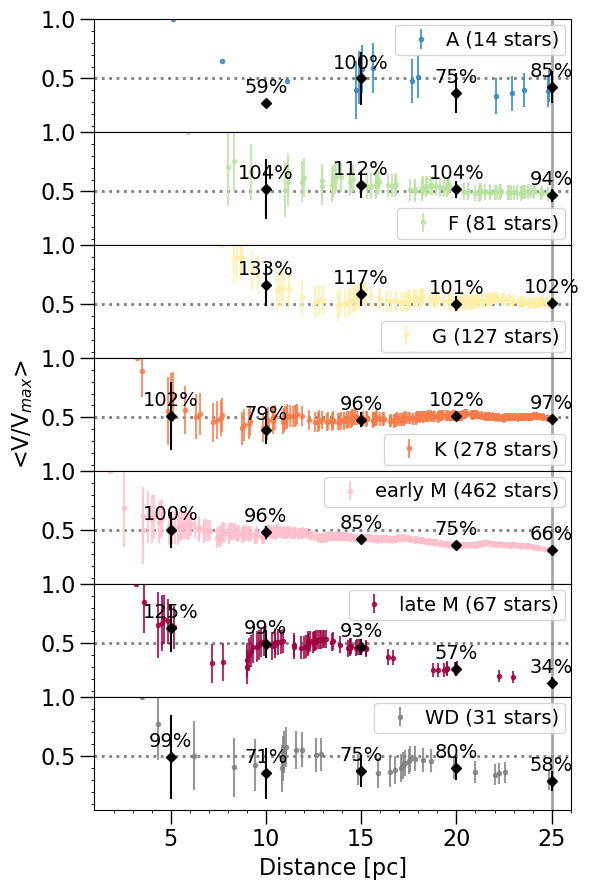}}
\caption{\textbf{(a):} $\langle V/V_{max} \rangle$ statistic to assess completeness of our volume-limited sample (\textit{top}). Percentages are the completeness of the sample out to the corresponding distance. Our sample is 90\% complete to 15~pc and 78\% complete to 25~pc. The green line traces the cumulative distance distribution of all observed stars in our sample (\textit{bottom}). \textbf{(b):} $\langle V/V_{max} \rangle$ statistic to assess completeness of our volume-limited sample, shown for each spectral type subsample.}
\label{fig:VVmax_dists}
\end{figure}

We evaluated the completeness of our volume-limited sample by examining its $\langle V/V_{max} \rangle$ statistic \citep{Schmidt68}. $\langle V/V_{max} \rangle$ is the mean of the distribution of $V/V_{max}$ values, which are the ratios of the volume out to the distance of each star to the maximum volume of the sample (i.e., spherical volume of radius = 25~pc). If a sample is 100\% complete, the mean of the $V/V_{max}$ values (which ranges from 0 to 1) should be $\langle V/V_{max} \rangle = 0.5$, meaning that the sample is uniformly distributed in space. A sample with an over-abundance of targets at greater distances within the volume will result in a $\langle V/V_{max} \rangle > 0.5$ and thus a completeness of $>$100\%. Figure \ref{fig:VVmax_dists}(a) shows the cumulative distribution of stellar distances in our sample and the $\langle V/V_{max} \rangle$ value of our sample as a function of corresponding limiting distance. At 25~pc, $\langle V/V_{max} \rangle = 0.39 \pm 0.01$, which corresponds to 78\% completeness. Our sample has 90\% completeness out to 15~pc. Figure \ref{fig:VVmax_dists}(b) shows the same $\langle V/V_{max} \rangle$ statistic for each spectral type subsample. F, G, and K stars are the most complete samples, with $\geq 94\%$ completeness at 25~pc, while M stars are less uniformly distributed as we go to larger distances. This is likely due to the difficulty in detecting M stars, which are less luminous and thus harder to detect, especially as the RECONS list was constructed before Gaia was available. Furthermore, we have a majority of early M type stars (M0--4) compared to late M type stars (M5 and later) due to the large decrease in luminosity across M type stars, making later M type stars fainter and harder to detect than early M type stars. There are too few A stars and white dwarfs to expect those samples to be complete.

\section{Data Reduction}

\label{sec:datareduction}
The data reduction process was automated as much as possible for efficiency and consistency. We followed the proven methods detailed in \citet{law14} and \citet{ziegler15}. An initial pipeline calibrates the images and performs the post-facto image displacement correction. The reduced images are then subject to a point-spread function (PSF) subtraction before searching for companions.

\subsection{Imaging Pipeline}
\label{sec:pipeline}

The Robo-AO imaging pipeline \citep{law09, law14} reduced the images: the raw EMCCD output frames are dark-subtracted and flat-fielded and then aligned and stacked using the Drizzle algorithm \citep{fruchter02}, which also increases the sampling in the images by a factor of two.  To avoid tip/tilt anisoplanatism effects, the image displacement was corrected during the reduction by using the target star itself as the guide star in each observation.

\subsection{PSF Subtraction}
\label{sec:psfsubtraction}

A custom PSF subtraction routine \citep{law14} based on the Locally Optimized Combination of Images algorithm \citep{lafreniere07} was applied to centered cutouts of all stars. Detailed in \citet{law14}, the code uses a set of twenty observations from the same observing night as reference PSFs, as it is improbable that a companion would appear at the same position in two different images.  A locally optimized PSF is generated and subtracted from the original image, leaving residuals consistent with photon noise.

\begin{figure}
\centering
\includegraphics[width=245pt]{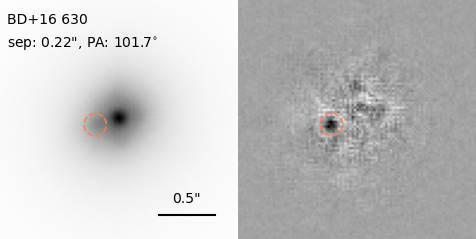}
\caption{Example of a companion detection in the original reduced image shown in linear-stretch (\textit{left}) and the PSF-subtracted image (\textit{right}). The companion is circled in orange.}
\label{fig:psf}
\end{figure}

This procedure was performed on all target star images out to 2$\arcsec$, and the results manually checked for companions. Figure$~\ref{fig:psf}$ shows an example of the PSF subtraction. The PSF subtracted images were subsequently processed through the automated companion-finding routine, as described in Section \ref{sec:compsearch}.

\subsection{Imaging Performance Metrics}
\label{sec:imageperf}

An automated routine was used to classify the image performance for each target.  Described in detail in \cite{law14}, the code uses PSF core size as a proxy for image performance, modeled by two Moffat functions to measure separately the widths of the core and halo of the PSF. Observations were binned into three performance groups by PSF core size, with 0.65\% falling in the low performance group (PSF core $<$ 0.10$\arcsec$), 9.83\% in the medium performance group (0.1 -- 0.15$\arcsec$), and 89.52\% in the high performance group ($>$ 0.15$\arcsec$).

\begin{figure}
\centering
\includegraphics[width=245pt]{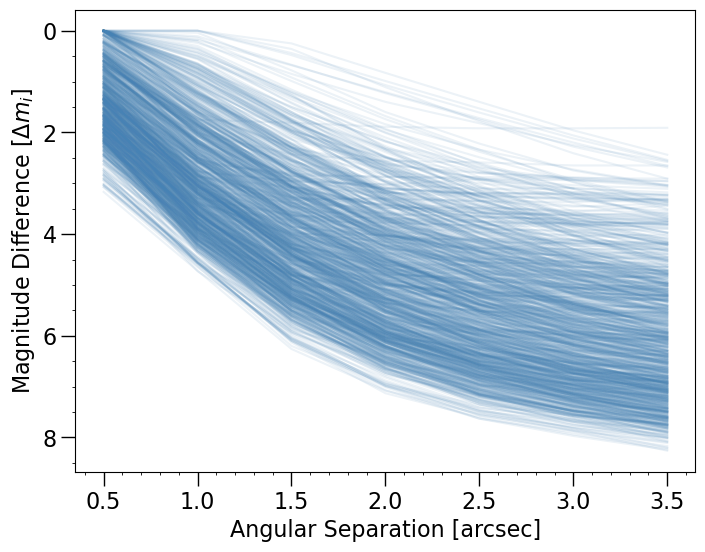}
\caption{Sensitivity curves for all observed targets.}
\label{fig:contrast_curves}
\end{figure}

We determine the sensitivity to a 5$\sigma$ detection by injecting artificial companions, created as a clone of the primary PSF at different angular separations and contrast ratios. For concentric annuli of 0$\farcs$1 width, the detection limit is calculated by steadily dimming the artificial companion until it falls below the 5$\sigma$ detection threshold for the auto-companion detection algorithm (Section \ref{sec:compsearch}). This process is subsequently performed at multiple random azimuths within each annulus and the limiting 5$\sigma$ magnitudes are averaged. For clarity, these average magnitudes for all radii measurements are fitted with functions of the form $a \times sinh(b \times r+c)+d$ (where \textit{r} is the radius from the target star and \textit{a, b, c} and \textit{d} are fitting variables). Achievable sensitivities are listed in Table~\ref{tab:roboaolist} and the sensitivity curves 
are shown in Figure$~\ref{fig:contrast_curves}$.

\section{Results}
\label{sec:results}

\subsection{Robo-AO Companion Detection}
\label{sec:compsearch}

\begin{figure*}
\centering
\includegraphics[width=500pt]{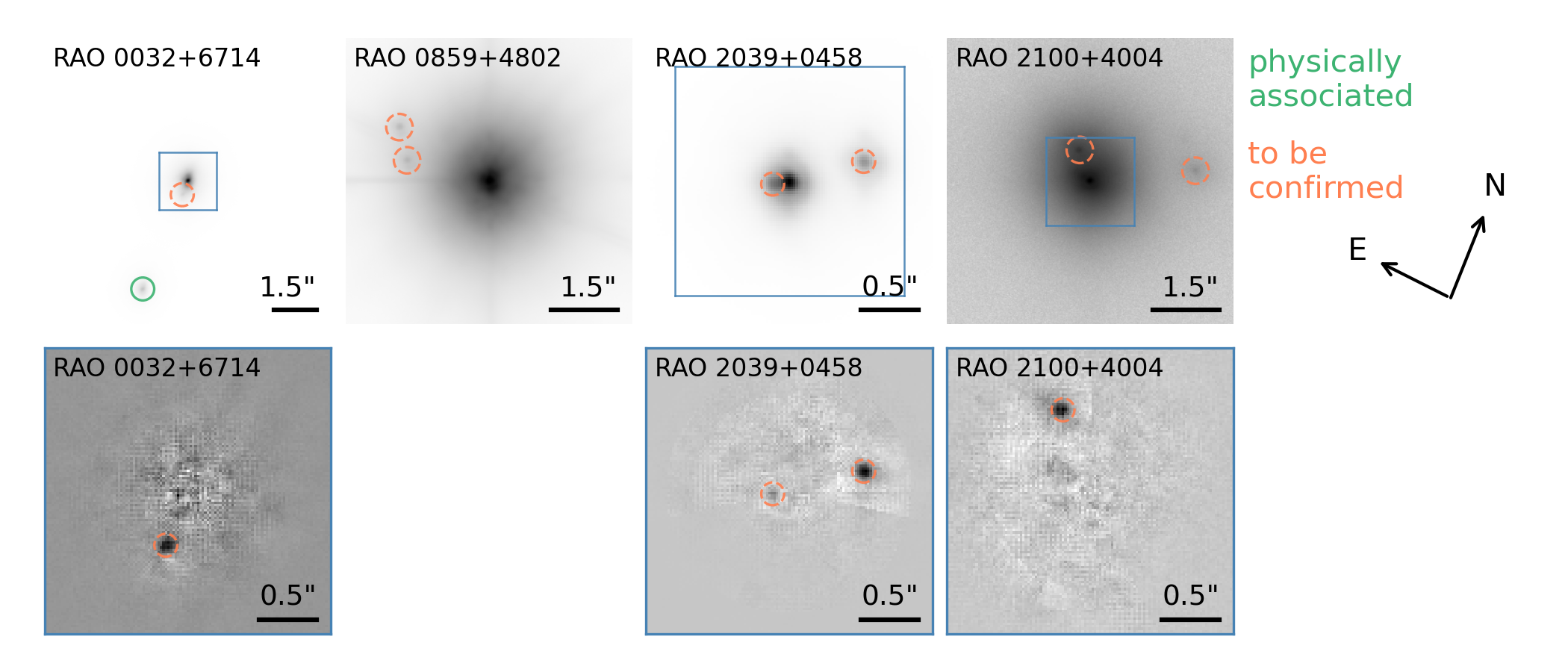}
\caption{Grid of Robo-AO images of tight triple system candidates. The images are displayed either in linear or log stretch (when $\Delta$m$>$2.5), for ease of visually seeing the fainter companion. The second row of images outlined in blue are the PSF-subtracted images in order to see the close-in companion(s) of the image above. The color of the circle indicating the companion is based on whether the companion is physically associated (solid green) or yet to be confirmed (dashed orange). The North-East orientation of the images is shown in the bottom right and the image scale in arcseconds is shown for each image cutout.}
\label{fig:images_triples}
\end{figure*}

To facilitate the automation of the data reduction, we created 8$\farcs$5 square cutouts centered on the observed target stars. Since we are primarily interested in those companions that can only be detected with high-angular resolution images (as opposed to seeing limited images), we selected a 4\arcsec separation cutoff for our companion search. We ran all images through a custom automated search algorithm, based on the code described in \citet{law14}. The algorithm slides a 5-pixel diameter aperture within concentric annuli centered on the target star. Any aperture with $>$5$\sigma$ outlier to the local noise is considered a potential astrophysical source. In addition, a visual companion search was performed redundantly by three of the authors. This search filtered out bad images, eliminated spurious detections with dissimilar PSFs to the target star and those having characteristics of a cosmic ray hit, such as a single bright pixel or bright streak. We detected \NcompsRAO\ tight companion candidates in Robo-AO images. The detection measurements and significance values are listed in Table~\ref{tab:comps_measurements} and the image cutouts are shown in Figures \ref{fig:images1}--\ref{fig:images_triples}.

\begin{figure*}
\centering
\includegraphics[width=500pt]{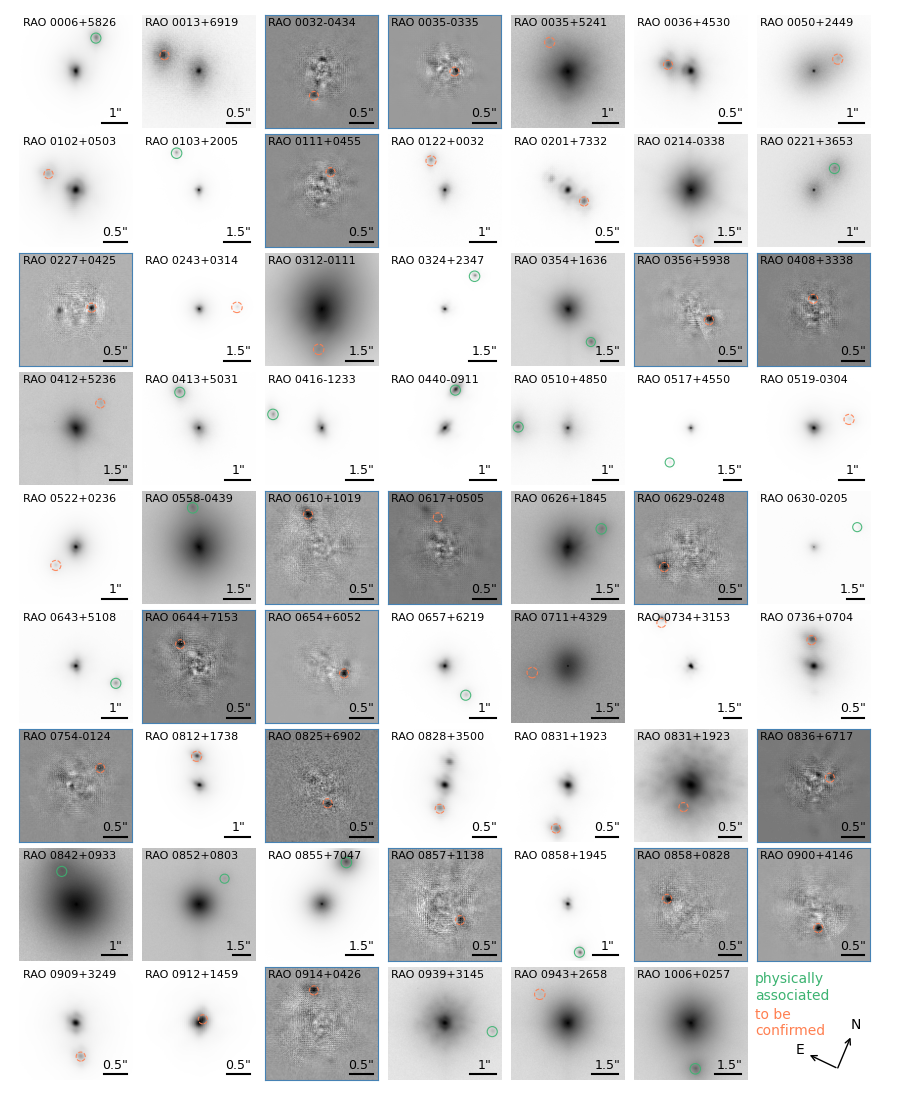}
\caption{Grid of Robo-AO images of binary candidates. The images are displayed either in linear or log stretch (when $\Delta$m$>$2.5), for ease of visually seeing the companion. The images outlined in blue are the PSF-subtracted images in order to see close-in companions. The color of the circle indicating the companion is based on whether the companion is physically associated (solid green) or yet to be confirmed (dashed orange). The North-East orientation of the images are shown in the bottom right and the image scale in arcseconds is shown for each image cutout.}
\label{fig:images1}
\end{figure*}

\begin{figure*}
\centering
\includegraphics[width=500pt]{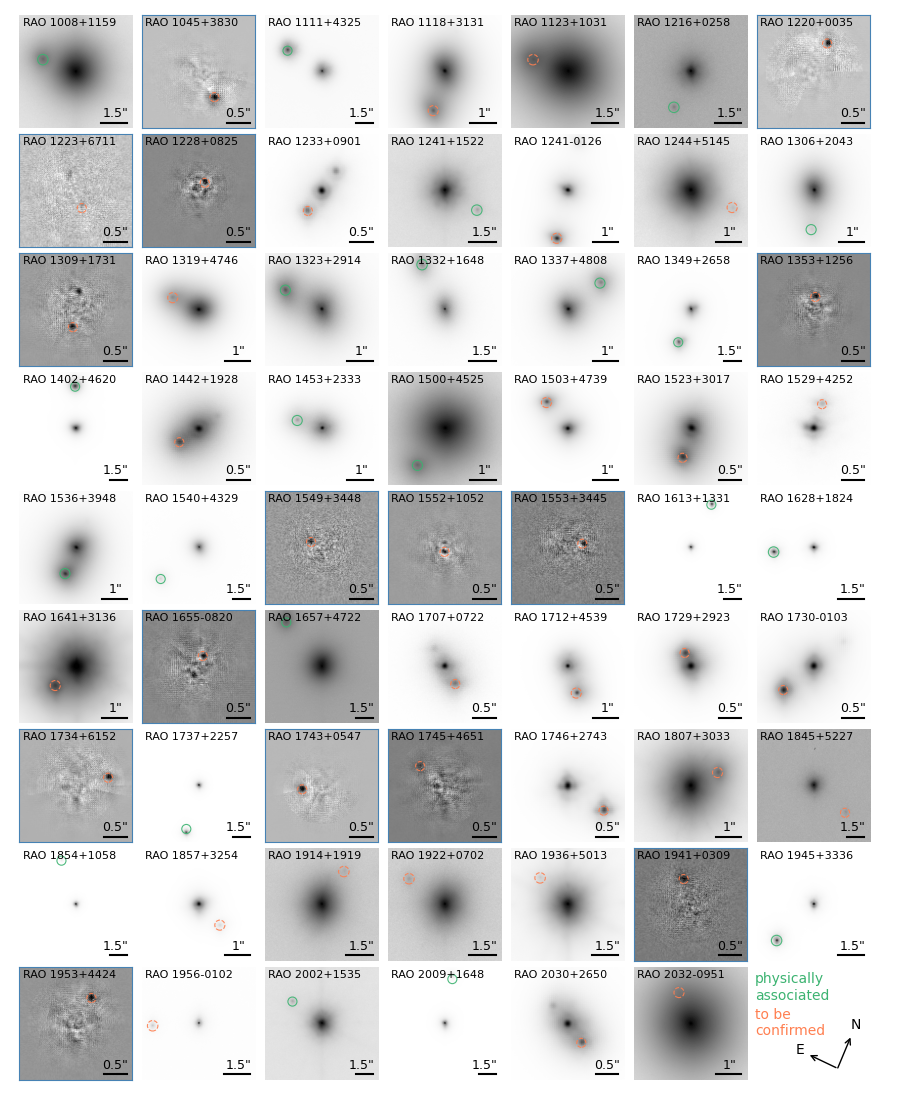}
\caption{(\textit{continued}). Grid of Robo-AO images of binary candidates. The images are displayed either in linear or log stretch (when $\Delta$m$>$2.5), for ease of visually seeing the companion. The images outlined in blue are the PSF-subtracted images in order to see close-in companions. The color of the circle indicating the companion is based on whether the companion is physically associated (solid green) or yet to be confirmed (dashed orange). The North-East orientation of the images are shown in the bottom right and the image scale in arcseconds is shown for each image cutout.}
\label{fig:images2}
\end{figure*}

\begin{figure*}
\centering
\includegraphics[width=500pt]{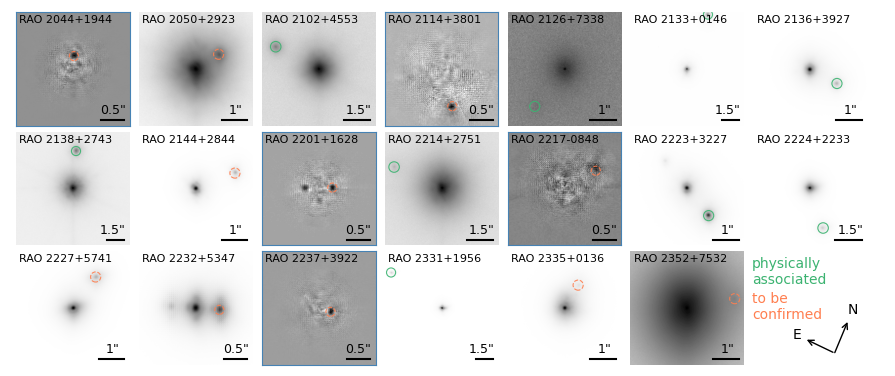}
\caption{(\textit{continued}). Grid of Robo-AO images of binary candidates. The images are displayed either in linear or log stretch (when $\Delta$m$>$2.5), for ease of visually seeing the companion. The images outlined in blue are the PSF-subtracted images in order to see close-in companions. The color of the circle indicating the companion is based on whether the companion is physically associated (solid green) or yet to be confirmed (dashed orange). The North-East orientation of the images are shown in the bottom right and the image scale in arcseconds is shown for each image cutout.}
\label{fig:images3}
\end{figure*}

\subsection{Companion Characterization}
\label{sec:characterization}

\subsubsection{Contrast Ratios}
\label{sec:contrastratios}

In order to measure the contrast ratio between the primary star and companion candidate, we removed most of the primary starlight by subtracting a radially averaged image. We then performed aperture photometry by measuring the flux in circular apertures around the primary star in the original image and the companion candidate in the radially-subtracted image. We measured the noise in an annulus around the star at the same separation as the companion candidate. The aperture radius was chosen to maximize the Signal-to-Noise ratio (S/N) of the companion detection. Photometric uncertainties were estimated from the standard deviation of the contrast ratios measured for the various aperture sizes. This method is explained in more detail in \cite{Salama21}.

\subsubsection{False Triples}
\label{sec:falsetriples}

Occasionally, the frame stacking centers on the bright companion instead of the target used as the natural guide star, which creates a ``false triple" image with the companion appearing on either side of the primary at the same separation and 180$^{\circ}$ rotated. We measured the contrast at both locations and combined the measurements following \cite{Law06} to calculate the final contrast. This affected 11 of our companion candidates. This effect can be seen in Figures \ref{fig:images1}-\ref{fig:images3} (e.g. RAO~\new{0201+7332}, RAO~\new{0828+3500}, RAO~\new{1233+0901}, ...)

\subsubsection{Separation and Position Angles}
\label{sec:separationposangles}

Raw pixel positions were calibrated to on-sky positions using a distortion solution produced using Robo-AO measurements for the globular cluster M15.\footnote{S. Hildebrandt (2013, private communication).} Separation and position angles were determined from the calibrated positions. Uncertainties were found using estimated systematic errors due to blending between the primary and close companions in the system. Typical uncertainty in the position for each star was 1-2 pixels \citep{Baranec16}, which corresponds to an average uncertainty of 0.06$\arcsec$. \new{For very tight companions ($\lesssim 0.5\arcsec$), a small uncertainty in companion position still results in a large position angle uncertainty ($\gtrsim 7^{\circ}$).}

\subsection{Gaia Companion Detections}

\subsubsection{Physical Association of Robo-AO Detections}
\label{sec:physicalassociation}

\begin{figure}
\centering
\includegraphics[width=245pt]{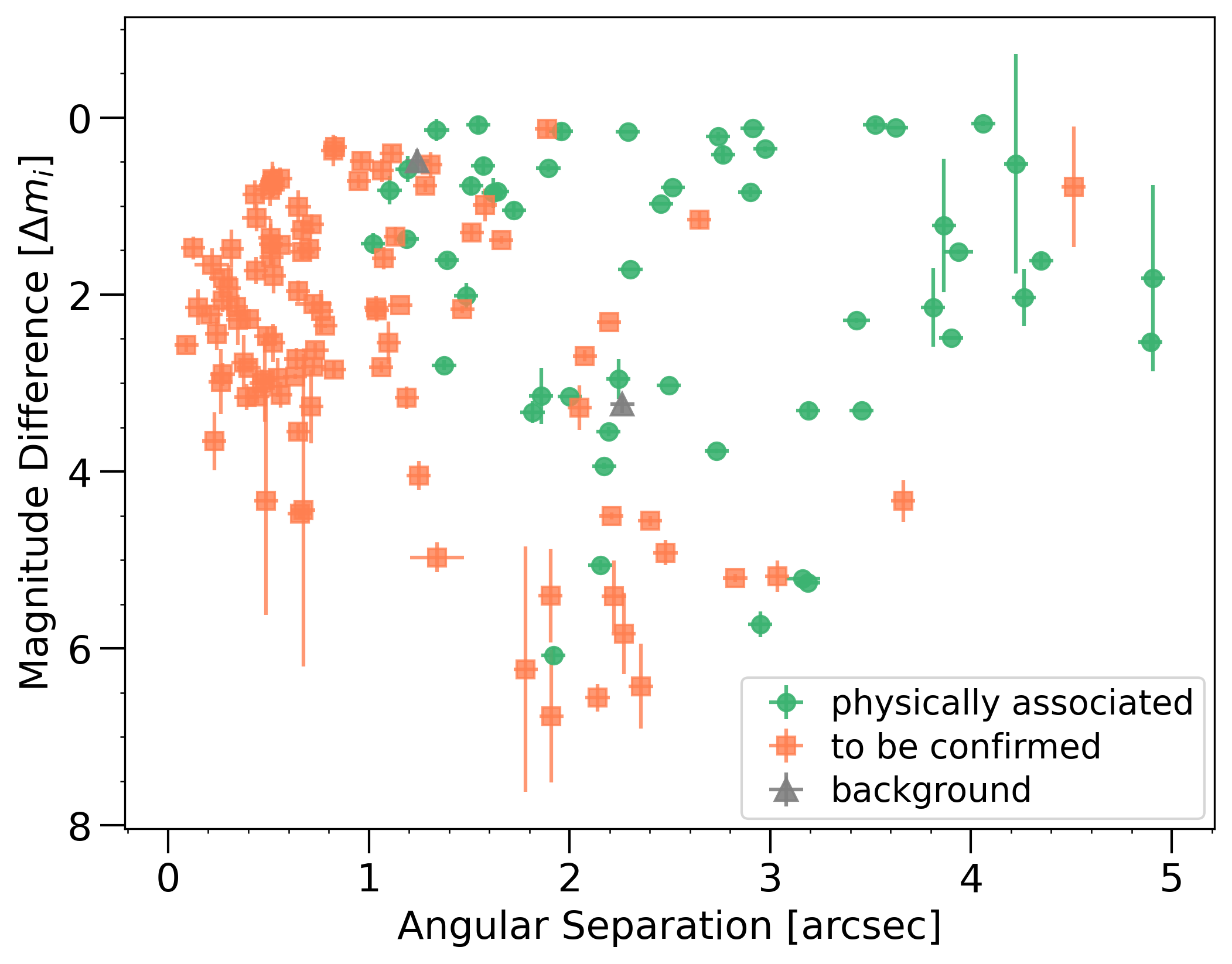}
\caption{Detected companion candidates within 4$\arcsec$ with Robo-AO. Companions with Gaia EDR3 parallax and proper motion measurements consistent with their host star are determined to be physically associated (green circles). Companion candidates with Gaia parallax and proper motion measurements inconsistent with the host star are determined to be background objects (grey triangles). The remaining companion candidates were not found in Gaia EDR3 or the star or companion were lacking parallax and proper motion information to confirm co-moving physical association (orange squares). }
\label{fig:comp_detections}
\end{figure}

\begin{figure}
\centering
\includegraphics[width=245pt]{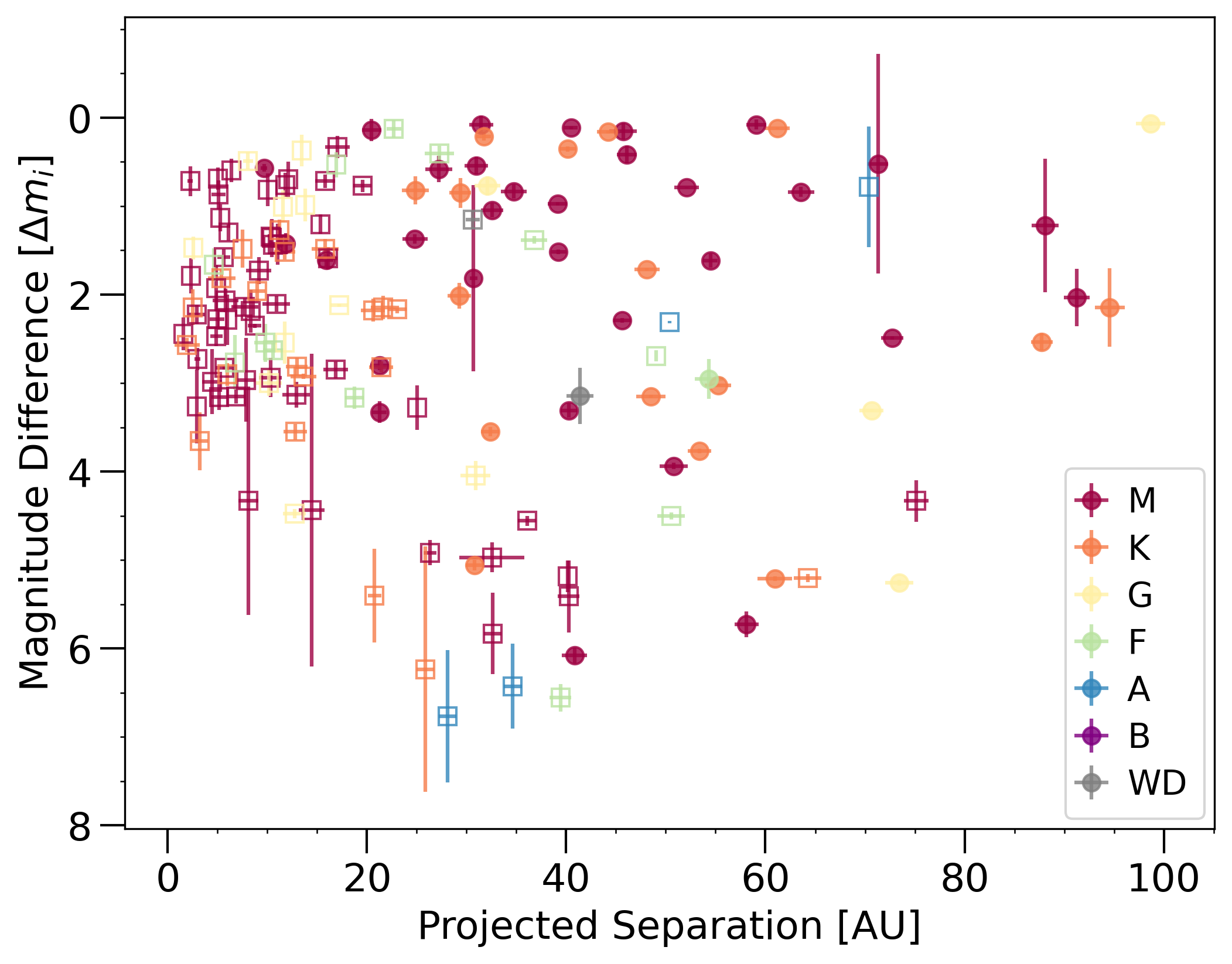}
\caption{Detected companions within 4$\arcsec$ as a function of projected separation in astronomical units and color-coded by primary star spectral type. Filled circles are physically associated companions and empty squares are yet to be confirmed companion candidates, where the distance was assumed to be the same as the host star. The detections determined to be background objects were removed.}
\label{fig:comp_detections_SpT}
\end{figure}

To determine if a candidate companion was physically associated with its host star, we searched for it in Gaia EDR3 to determine whether it had matching parallax and proper motion measurements. To determine physical association status, we applied a threshold of 0.35 to both the ratio of the difference in parallax measurements between the primary and companion to the primary's parallax ($\Delta \pi / \pi_{prim} < 0.35 $) and to the ratio of the difference in proper motion measurements between the primary and companion to the primary's proper motion ($\Delta \mu/\mu_{prim} < 0.35 $), as explained in \cite{Salama21}. We determined \NcompsboundRAO\ companions detected with Robo-AO to be physically associated and ruled out \Ncompsbckgrd\ as background objects. Another \Ncompsunknown\ were not found in Gaia EDR3 or were lacking parallax and proper motion measurements and thus remain to be confirmed with future follow-up studies. However, we expect the vast majority of these companion candidates to be physically associated, as they are mostly within 1$\arcsec$, where the likelihood of a background star alignment is much lower \citep{Horch14,Ziegler20,Salama21}. The \textit{i}\textsuperscript{$\prime$}-band contrasts of the \NcompsRAO\ companion candidates detected with Robo-AO are shown in Figure~\ref{fig:comp_detections} as a function of separation from the host star with their physical association status denoted. Figure~\ref{fig:comp_detections_SpT} also shows the companion detections, as a function of projected separations in AU and the primary star's SpT is shown.

\subsubsection{Wide Co-moving Companions}
\label{sec:GaiaComove}

\begin{figure}
\centering
\includegraphics[width=245pt]{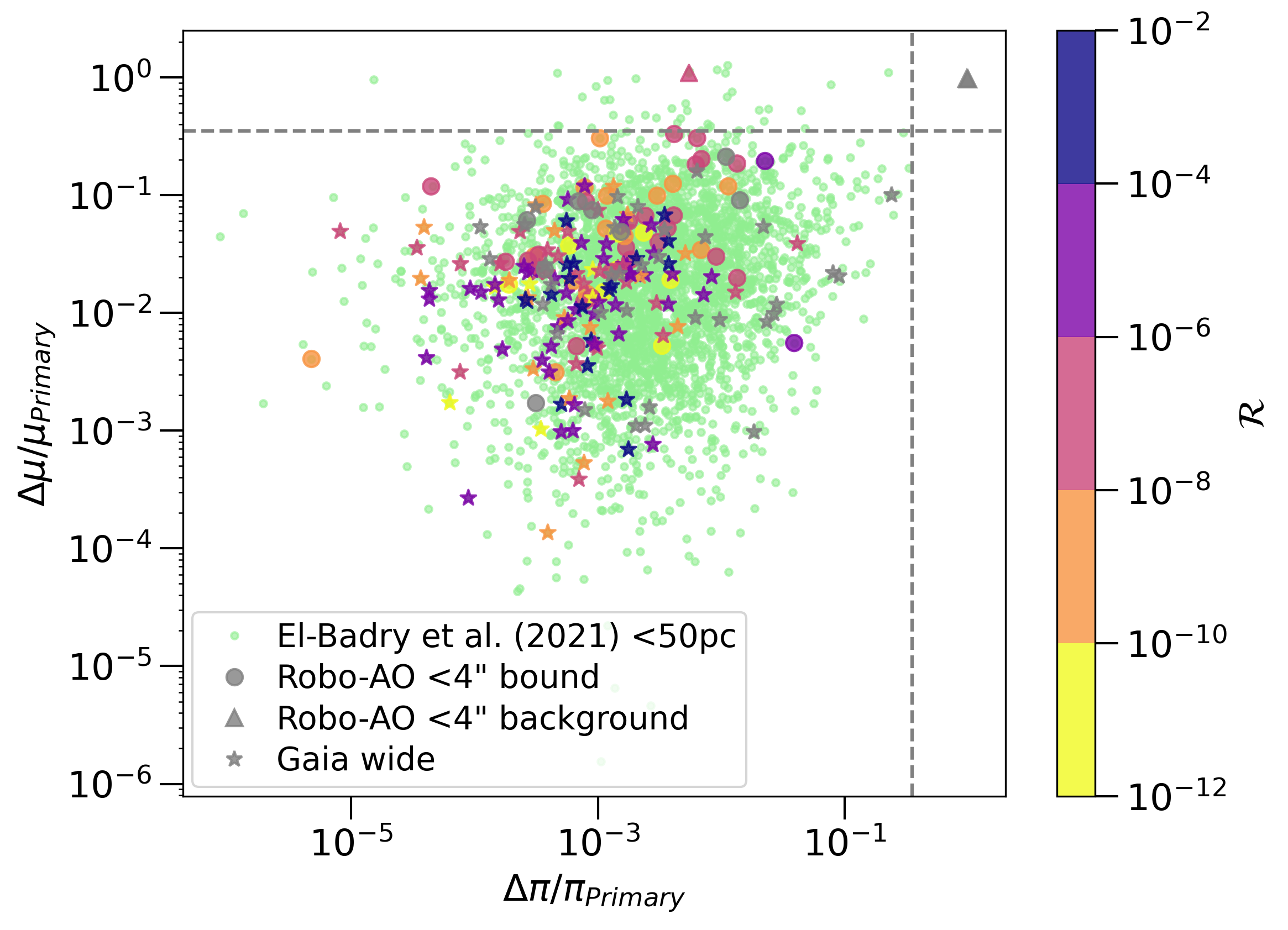}
\caption{Physical association thresholds (the two dashed lines at 0.35), identified by comparing the primary and companion proper motion and parallax Gaia EDR3 measurements. The wide Gaia co-moving companions as well as those detected in Robo-AO images with Gaia EDR3 information are shown. For those found in the \cite{ElBadry21} catalog, their $\mathcal{R}$ value is given and we show the sub-sample of stars within 50pc in \cite{ElBadry21}'s binary catalog in green.}
\label{fig:comps_wide_physassoc}
\end{figure}

\begin{figure}
\centering
\includegraphics[width=245pt]{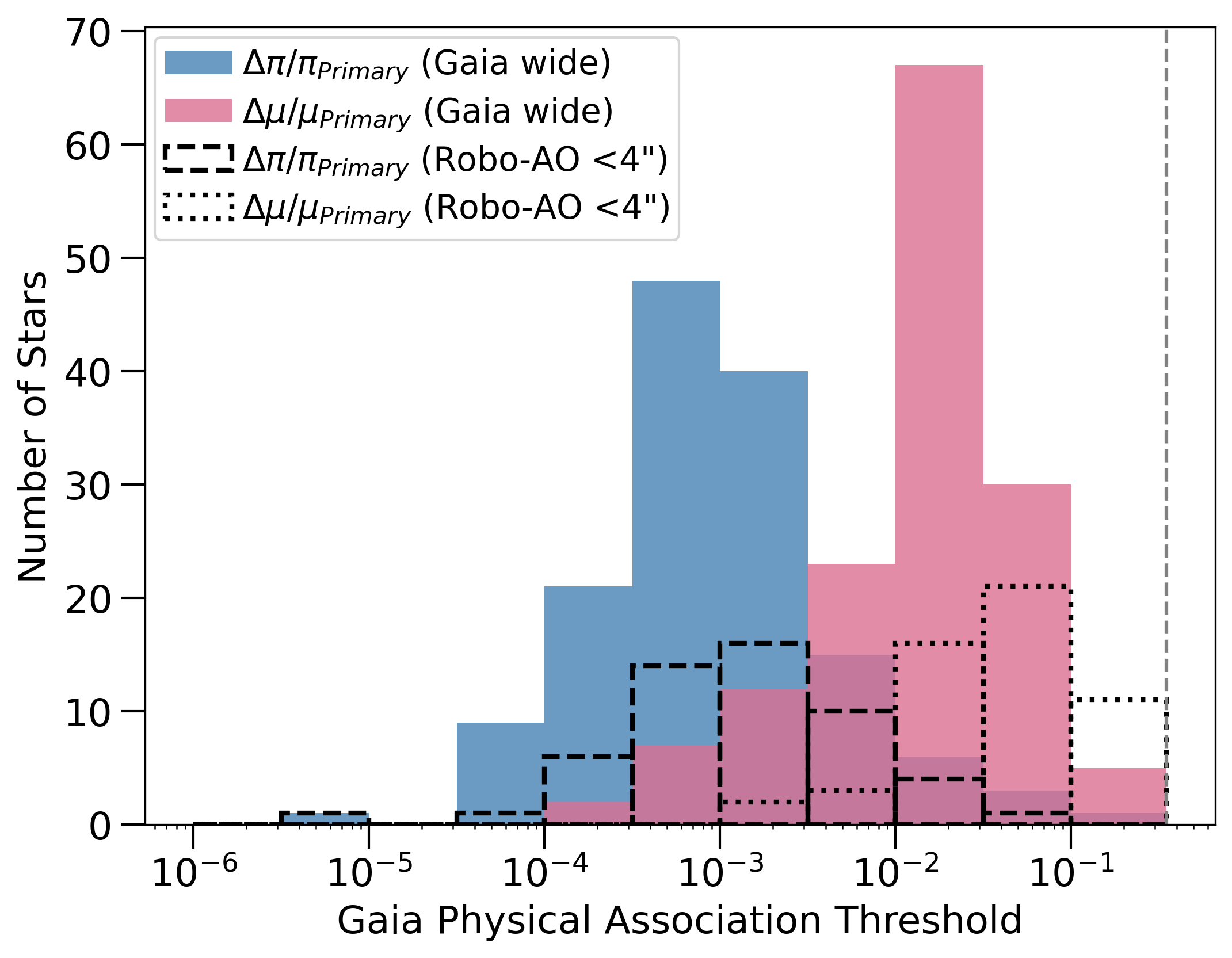}
\caption{The distribution of $\Delta \pi / \pi_{prim}$ and $\Delta \mu / \mu_{prim}$ for the wide co-moving companions identified in Gaia EDR3 as well as the Robo-AO-detected companions found in Gaia EDR3.}
\label{fig:physassoc_hist}
\end{figure}

In order to evaluate the multiplicity fractions of our sample, in addition to the close companions detected with Robo-AO within 4$\arcsec$, where AO is most beneficial, we searched for wide co-moving companions out to separations of 10,000~AU using Gaia EDR3. We found an additional \NcompsGaiaDet\ wide companions. This wide co-moving companion search was carried out for the \NstarsGaia\ stars found in Gaia EDR3. We used the same $\Delta \pi / \pi_{prim} < 0.35 $ and $\Delta \mu/\mu_{prim} < 0.35 $ thresholds as described in Section \ref{sec:physicalassociation} to determine co-moving companions. As can be seen in Figures \ref{fig:comps_wide_physassoc} \& \ref{fig:physassoc_hist}, all but one companion have $\Delta \pi / \pi_{prim} < 0.1 $ and 97\% have $\Delta \mu / \mu_{prim} < 0.1 $. 

To further validate our companion search and physical association analysis, we cross-matched with \cite{ElBadry21}'s catalog of one million binaries in Gaia EDR3. The colored symbols in Figure \ref{fig:comps_wide_physassoc} are the companions that are found in \cite{ElBadry21}'s catalog and thus have $\mathcal{R}$ values, which can be used as a proxy for the probability that a companion candidate is instead a chance alignment. $\mathcal{R}$ is computed from the ratio of the estimated density of chance alignments at a given point in parameter space to the estimated density of candidates (which includes both true companions and chance alignments): $\mathcal{R}(x) = \mathcal{N}_{chance ~align}(x)/\mathcal{N}_{candidates}(x)$. A value of $\mathcal{R}$ very close to 0 means a very high probability that the companion candidate is a true physically associated object, while a value of $\mathcal{R}$ close to 1 indicates a high probability that the object is a chance alignment. The highest $\mathcal{R}$ value in our sample of companions found in \cite{ElBadry21}'s catalog is 0.0068, therefore we determine them all to be physically associated. Furthermore, for those not found in the catalog and that overlap in $\Delta \pi / \pi_{prim}$ and $\Delta \mu / \mu_{prim}$ space with those from the catalog, we determine them to also be physically associated.

Therefore, in total \Ncompsbound\ companions are determined to be physically associated: \NcompsboundRAO\ detected with Robo-AO and \NcompsGaiaDet\ from the Gaia wide co-moving search (\NcompsGaiawide\ as new companions and \NcompsRAOprims\ are wide companions that were observed separately from the primary star with Robo-AO allowing a search for tight tertiary companions around the secondary star). Using \cite{ElBadry21}'s catalog, we further validate \NcompsElBadry\ of the \Ncompsbound\ companions as physically associated.

\subsubsection{Companion Properties}
\label{sec:GaiaProp}

\begin{figure}
\centering
\includegraphics[width=245pt]{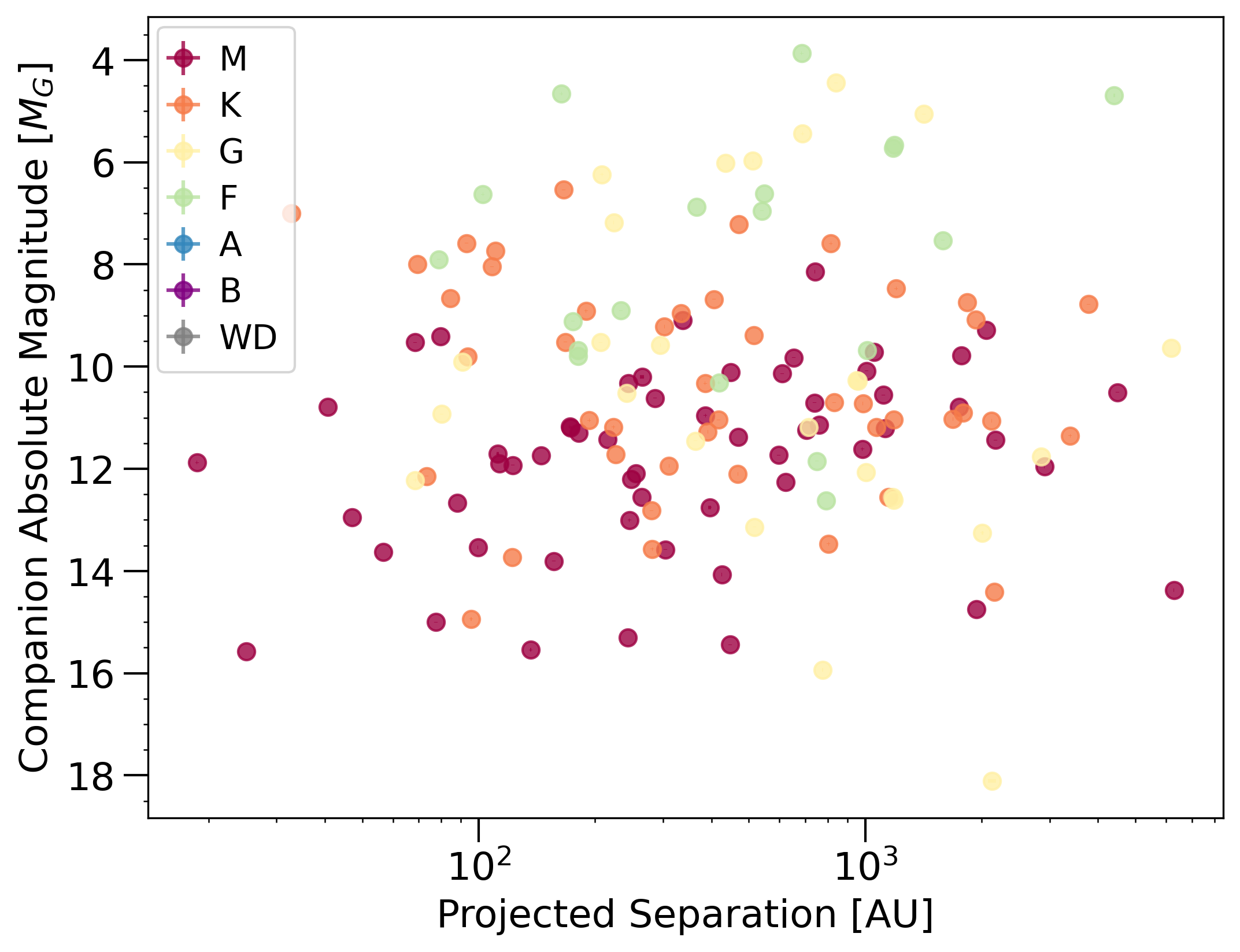}
\caption{Absolute Gaia magnitudes of \NcompsGaiaDet\ wide co-moving companions identified in Gaia EDR3 as a function of projected separation and color-coded by spectral type of the primary star.}
\label{fig:comps_wide_AU}
\end{figure}

The detected co-moving companion absolute Gaia magnitudes as a function of projected separation are shown in Figure \ref{fig:comps_wide_AU}. The farthest detected companion is at a projected separation of 6285~AU. There are three companions found in the Gaia co-moving search that are within 4$\arcsec$ but missed by Robo-AO: G~239-25 was near the edge of the detector in the observation and thus the companion was out of the field of view; 16~Cyg~A has a companion with too large contrast ($\Delta m_G \approx 8$); and VV~Lyn~A has too poor image quality for Robo-AO to detect the companion. 

\begin{figure}
\centering
\includegraphics[width=245pt]{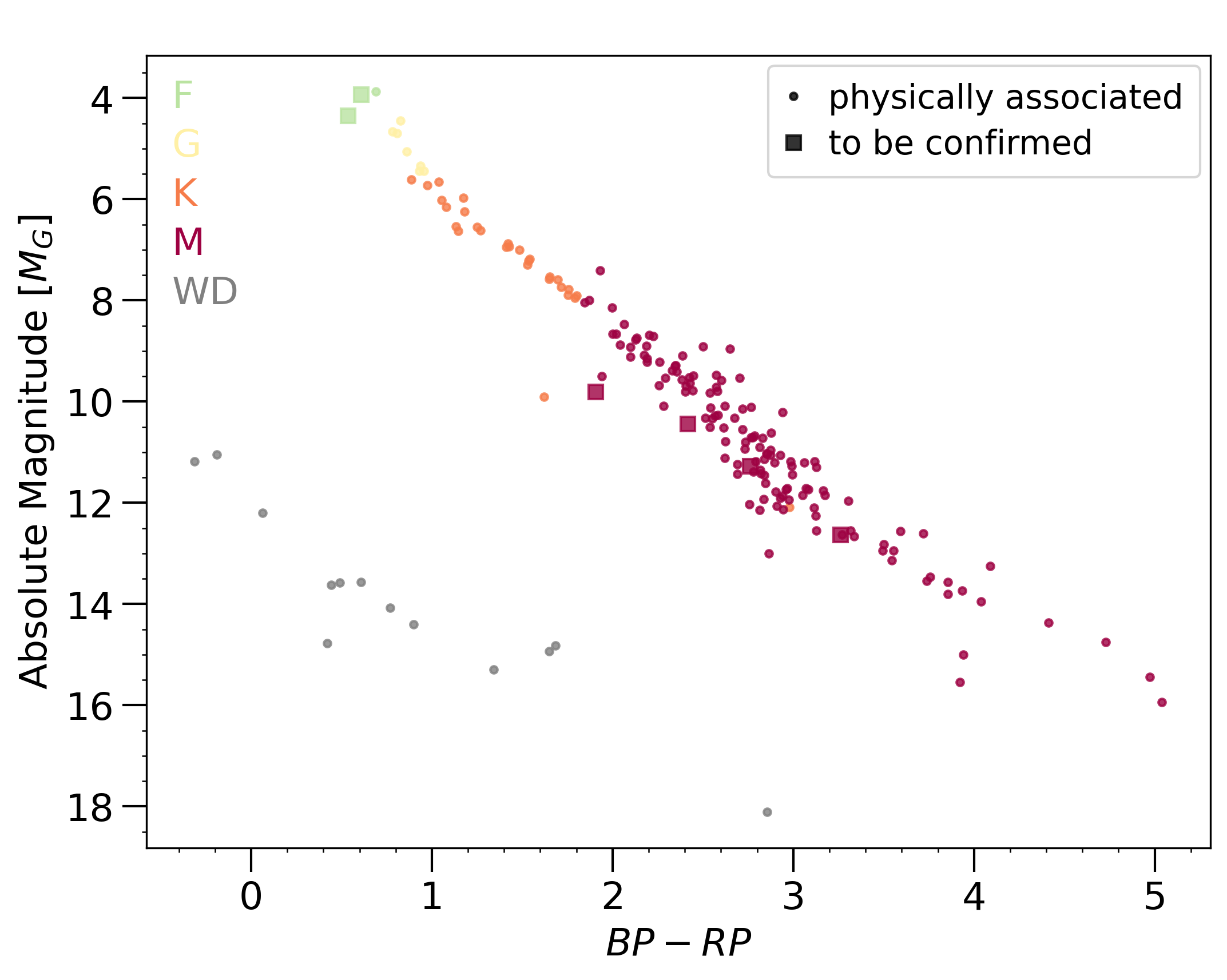}
\caption{Gaia color-magnitude diagram for companions detected in Gaia, including the wide co-moving search and the Robo-AO tight detections found in Gaia. Color-coded by companion SpT estimate from Gaia photometry.}
\label{fig:comps_CMD}
\end{figure}

Figure \ref{fig:comps_CMD} shows companion properties from Gaia magnitude and colors and color-coded by SpT estimate as explained in Section \ref{sec:targetverification}, this time for the companions found in Gaia EDR3. The \cite{Pecaut13} grid is only for main sequence stars, so companions that clearly fall in the white dwarf portion of the color-magnitude diagram were determined to be white dwarfs from their placement on the plot rather than directly from the \cite{Pecaut13} grid. This was also further validated for companions found in the \cite{ElBadry21} catalog, which labeled all binaries with a white dwarf component.

\subsection{Cross-matching with other catalogs}
\label{sec:catalogs}

Of the \Ncomps\ total companions (including both the \NcompsRAO\ detected by Robo-AO and the \NcompsGaiaDet\ wide co-moving companions found with Gaia) reported in this survey, we find \NcompsCats\ of them in the literature and the remaining \NcompsNew\ are new discoveries. We found \NcompsRag\ of our companions around FGK stars reported in \cite{Raghavan10} and \NcompsWinters\ of those around M type stars reported in \cite{Winters19}. We checked for the remaining companions in the Washington Double Star catalog (WDS; \citealt{Mason01}) and found another \NcompsWDS\ and an additional \NcompsIntFourth\ in the Fourth Catalog of Interferometric Measurements of Binary Stars \citep{Hartkopf01}. They are all marked in Table \ref{tab:comps_measurements}. We also searched for our companions in \cite{Vrijmoet20}'s study where they calculated orbits for resolved multiple systems and flagged systems as potential multiples in a RECONS sample of 210 (mostly southern) M dwarfs. \NstarsRECOVrij\ of our observed stars are in their sample, including \NcompsRECOVrij\ of them where we detect companions with Robo-AO. Of these detected companions, \NcompsRECOVrijunr\ are either flagged as potentially having unresolved companions or marked as single stars. Another \NnocompsRECOVrijunr\ were marked as potentially having companions but we do not detect any in our Robo-AO images.

\section{Analysis \& Discussion}
\label{sec:discussion}

\subsection{Gaia companion recovery}

\begin{figure}
\centering
\includegraphics[width=245pt]{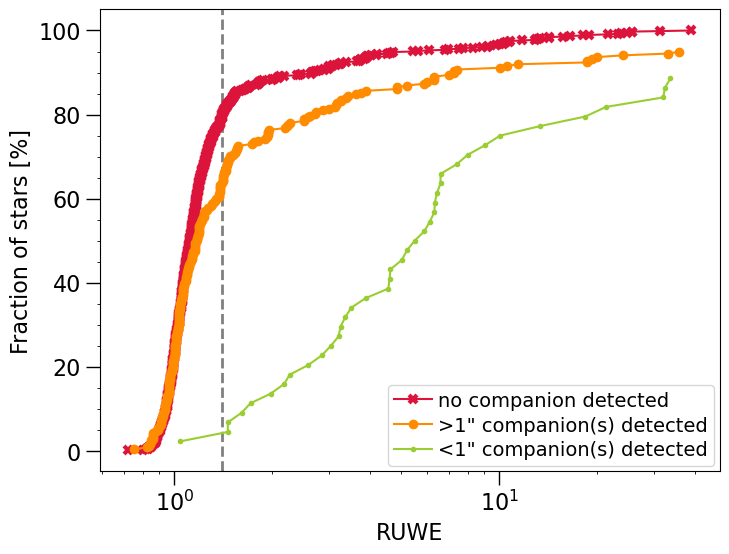}
\caption{Cumulative distribution of RUWE values for stars with companions detected by Robo-AO within 1$\arcsec$, stars with companion beyond $>$1$\arcsec$) only, and stars with no companions detected. The vertical dashed line marks RUWE = 1.4. Large RUWE values ($>$1.4) indicate an issue with the astrometry fitting to a single-star model, likely caused by the presence of a close companion.}
\label{fig:comp_cumul_RUWEs}
\end{figure}

\begin{figure}
\centering
\includegraphics[width=245pt]{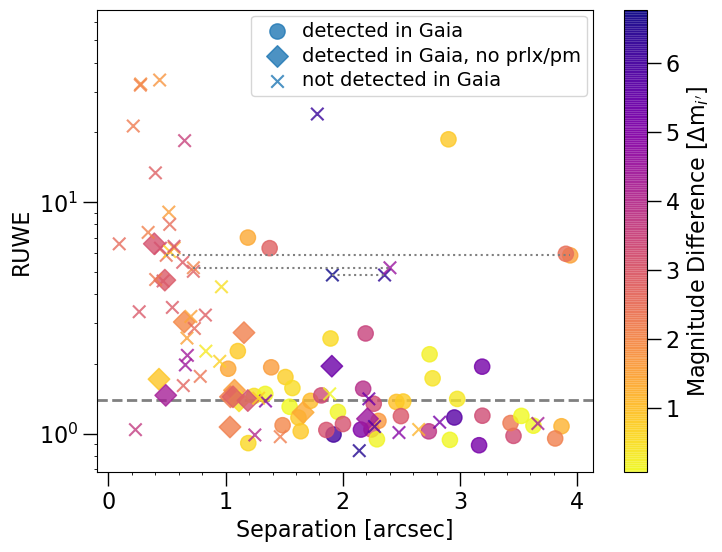}
\caption{Primary star RUWE values as a function of companion separations detected by Robo-AO, color-coded by magnitude difference. Companions that are detected by Gaia and have parallax and proper motion measurements are shown as circles, those that are found in Gaia but with no parallax and proper motion measurements are shown as diamonds, while those not found in Gaia are shown as X's. Companions that orbit the same primary star and detected by Robo-AO within 4$\arcsec$ are connected by a dotted line. The horizontal dashed line marks RUWE = 1.4. It is clear that companions within 1$\arcsec$ are not well detected by Gaia and are causing large RUWE values, while the systems where the companions are well detected generally have smaller RUWE values.}
\label{fig:comp_seps_RUWEs}
\end{figure}

Gaia Renormalized Unit Weight Errors (RUWE) values are often used as an indicator of the presence of a companion, as a large ($>$1.4) RUWE value indicates an issue with the single-star astrometric model (as seen in  \citealt{Ziegler20}). Figure \ref{fig:comp_cumul_RUWEs} shows the cumulative distribution of primary star RUWE values for our sample in three categories: systems with tight companions within 1$\arcsec$ detected by Robo-AO, systems with only companions beyond 1$\arcsec$, and single stars. For systems with companion detections, Figure \ref{fig:comp_seps_RUWEs} shows the correlation between primary star RUWE value and companion separation and whether a companion is detected by Gaia is also denoted. Gaia EDR3 does not resolve most companions detected in Robo-AO images within 1$\arcsec$, however the presence of a companion can be inferred from the high RUWE values.

\subsection{Accelerating Stars}
\label{sec:acceleratingstars}

\begin{figure}
\centering
\includegraphics[width=245pt]{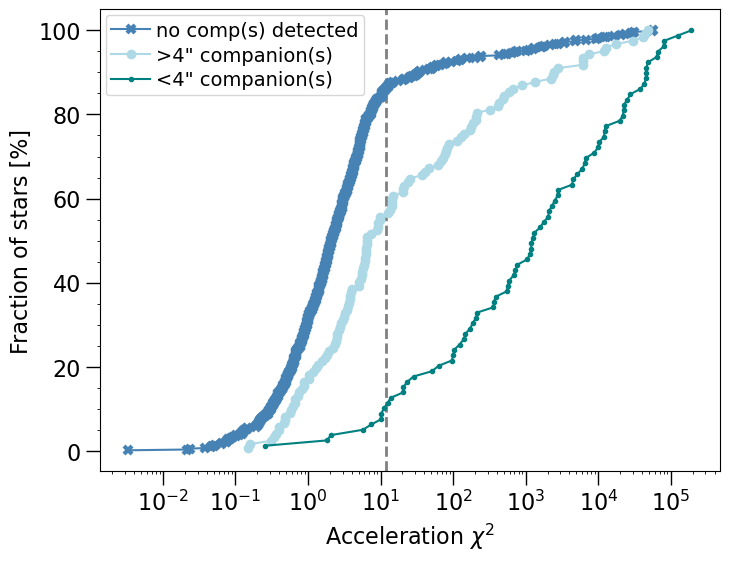}
\caption{Cumulative distribution of HGCA acceleration $\chi^2$ values for stars with and without companions detected. The vertical dashed line marks the threshold for significant acceleration ($\chi^2 > 11.8$). Of the \NstarsAccNOComps\ stars with no companions detected, \FracAccNOCompsNOsig\ (\NstarsAccNOCompsNOsig/\NstarsAccNOComps) do not have significant accelerations, while \FracAccNOCompsSig\ (\NstarsAccNOCompsSig/\NstarsAccNOComps) of them do. Of the \NstarsAccWideComps\ stars with wide ($>$4$\arcsec$) companion detections, \FracAccWideCompsNOsig\ (\NstarsAccWideCompsNOsig/\NstarsAccWideComps) do not have significant accelerations, and \FracAccWideCompsSig\ (\NstarsAccWideCompsSig/\NstarsAccWideComps) of them do. Of the \NstarsAccTightComps\ stars with companions detected within 4$\arcsec$, only \FracAccTightCompsNOSig\ (\NstarsAccTightCompsNOsig/\NstarsAccTightComps) do not have significant accelerations while \FracAccTightCompsSig\ (\NstarsAccTightCompsSig/\NstarsAccTightComps) of them do.}
\label{fig:acc_chisqs}
\end{figure}

\begin{figure}
\centering
\includegraphics[width=245pt]{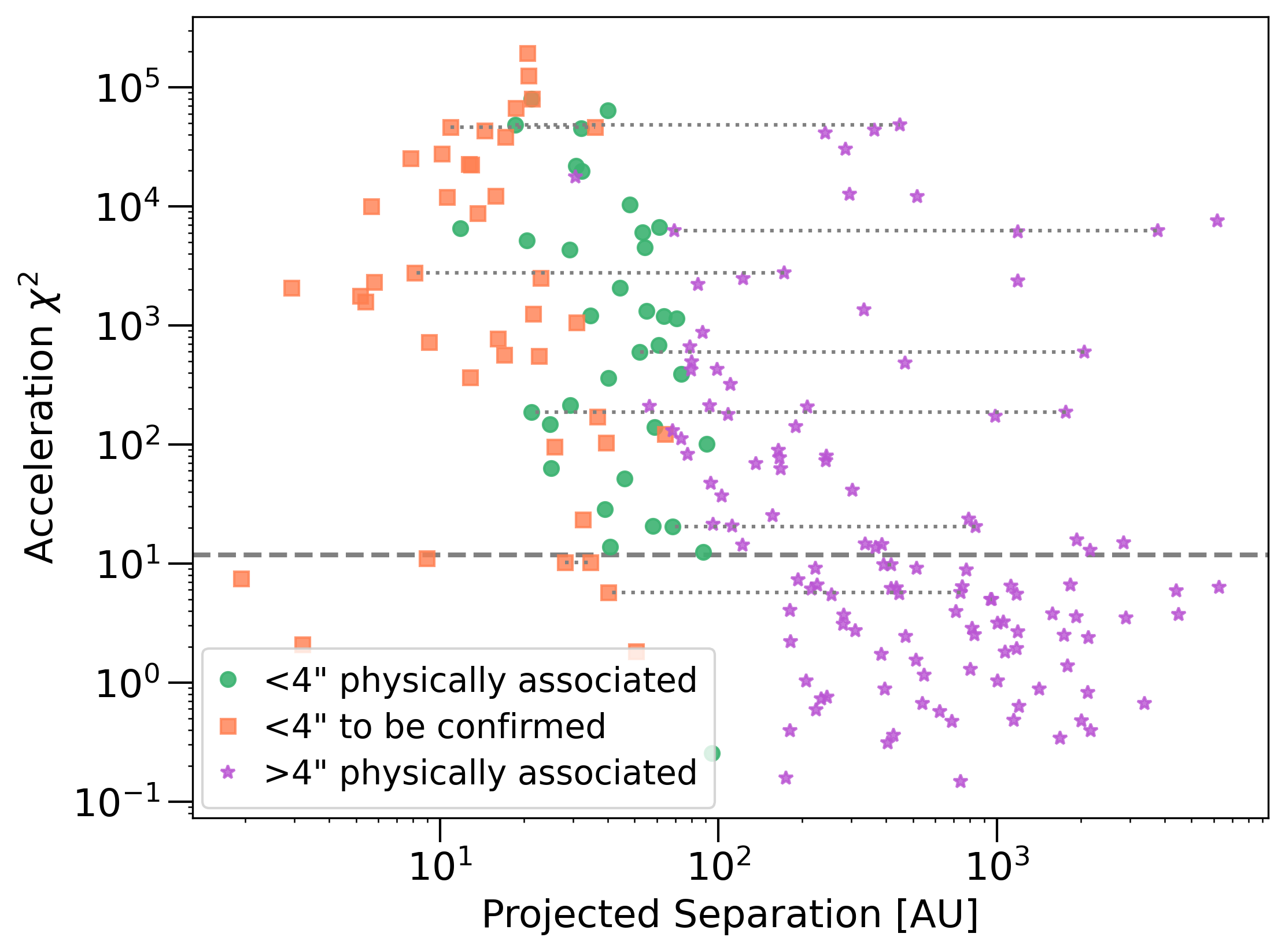}
\caption{Primary star HGCA acceleration $\chi^2$ as a function of companion projected separation. Orange squares are detections within 4$\arcsec$ with Robo-AO that have yet to be confirmed as physically associated, green circles are physically associated companions detected within 4$\arcsec$ with Robo-AO, and purple stars are wide co-moving companions detected in Gaia EDR3. Companions linked by a dotted line have the same primary star. The horizontal dashed line marks the threshold for significant acceleration ($\chi^2 > 11.8$).}
\label{fig:acc_chisqs_seps}
\end{figure}

Dynamical masses and orbital information on companions can be determined from changes in stellar proper motion. The Hipparcos-Gaia Catalog of Accelerations (HGCA; \citealt{Brandt18,Brandt21}) comprises stellar accelerations determined from Hipparcos and Gaia EDR3 proper motion and positional measurements. Stars with acceleration measurements of $\chi^2 > 11.8$ (corresponding to 3$\sigma$) are considered to have significant accelerations, likely caused by the presence of a companion. Of the \NstarsAcc\ stars from our sample that are found in the HGCA, \NstarsAccSig\ have significant accelerations. We report companions within 4$\arcsec$ detected with Robo-AO around \NstarsAccTightCompsSig\ of those stars, with another \NstarsAccWideCompsSig\ stars having wide co-moving companions found in Gaia EDR3. We also detected companions around \NstarsAccCompsNOsig\ of the \NstarsAccNOsig\ stars with no significant acceleration measurements (\NstarsAccTightCompsNOsig\ within 4$\arcsec$ detected by Robo-AO and \NstarsAccWideCompsNOsig\ wide co-moving companions found in Gaia EDR3). Figure~\ref{fig:acc_chisqs} shows the cumulative distribution of acceleration $\chi^2$ for stars with no companions detected, with only wide companions beyond 4$\arcsec$, and with any companions within 4$\arcsec$. Figure~\ref{fig:acc_chisqs_seps} shows the correlation between significance of acceleration and companion separation. As can be seen in Figure \ref{fig:acc_chisqs}, the distribution that consists of wide companions has a majority of stars with non-significant accelerations (\FracAccWideCompsNOsig), even though they have companions. However, the distribution only including the tight companions, has a much larger fraction (\FracAccTightCompsSig) with significant accelerations. This is expected, as companions at larger separations will have less effect on the primary star's acceleration. In the sample with no detected companions, whether tight or wide, there are still some stars with significant accelerations (\FracAccNOCompsSig). It is likely that these stars have unresolved companions.

\subsection{Higher-Order Systems}
\label{sec:highorders}

Triple systems and higher-order systems give us important insights on the dynamics and formation histories of stellar systems. The only stable orbital configuration for these systems is a hierarchical structure, with a tight binary and wider tertiary \citep{Reipurth12}. 

We detected \NhigherOrdersTight\ triple systems with both companions within 4$\arcsec$. They are shown in Figure \ref{fig:images_triples}. The hierarchical structure can clearly be seen for three of the four systems. With our wide co-moving companion search in Gaia EDR3, we found an additional 17 higher-order systems. \NhigherOrdersBoth\ are combinations of a tight binary (detected with Robo-AO) and a wider co-moving tertiary companion identified in Gaia, and the other \NhigherOrdersWide\ have both companions as wide co-moving companions identified with Gaia. In total, we detected \Ntriples\ triple systems and \Nquads\ higher-order systems, detailed below. 

\begin{figure}
\centering
\includegraphics[width=245pt]{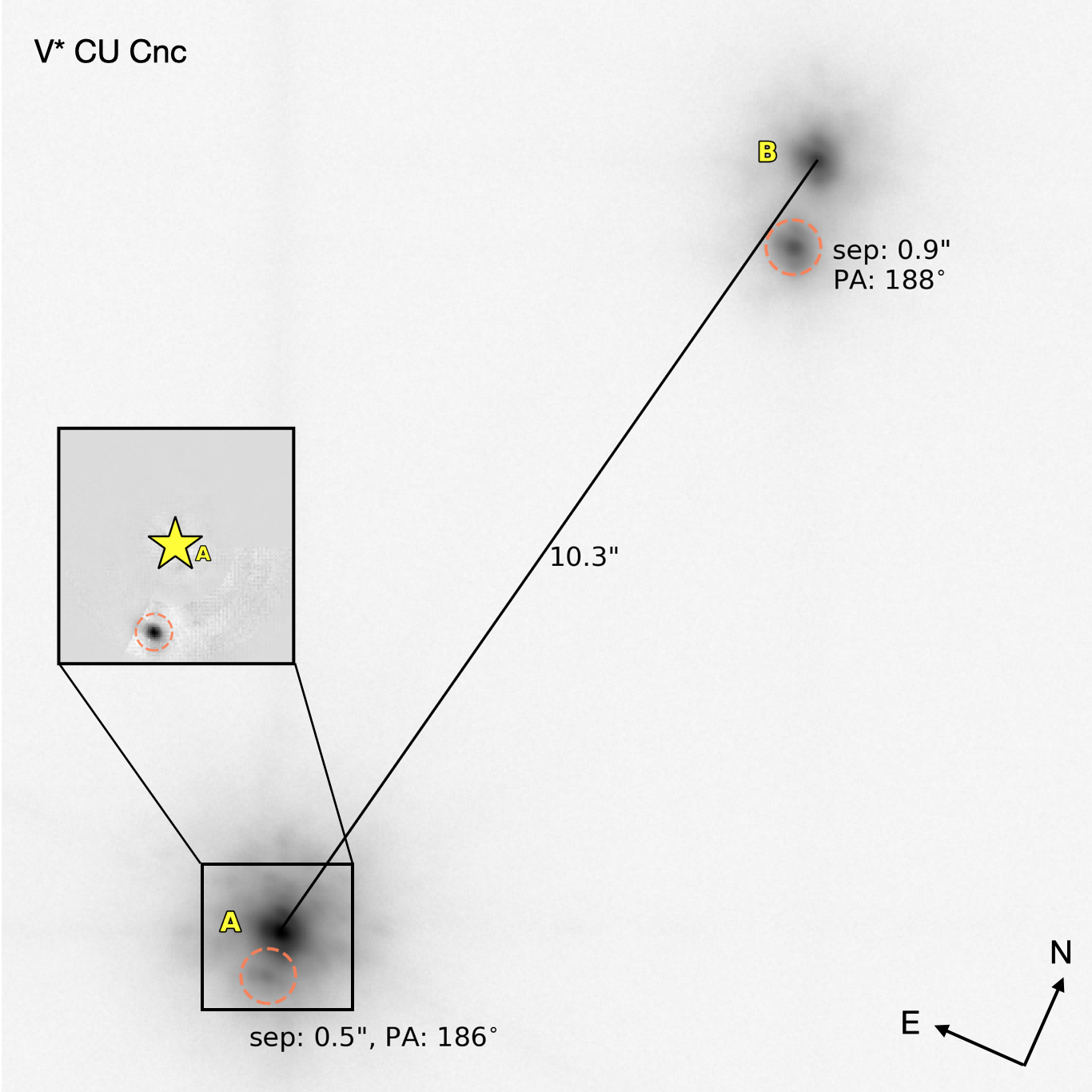}
\caption{CU CnC system consisting of four imaged M dwarfs with the PSF-subtracted image of star A in order to see its companion. The Robo-AO image is displayed with a log stretch.}
\label{fig:quadruple}
\end{figure}

We resolved 4 of the 5 known components of the Cu CnC quintuple system (with the 5th being an eclipsing binary to primary star Cu CnC A). The imaged system consists of two sets of tight ($<1\arcsec$) binaries separated by $\approx$10$\arcsec$ (Figure \ref{fig:quadruple}). \cite{Wilson17} recently studied this system and \cite{Beuzit04} imaged it with adaptive optics in 2000, reporting separations and position angles of 0.68$\arcsec$ and 158$^{\circ}$ for Cu CnC AD and 0.55$\arcsec$ and 219$^{\circ}$ for BC. Twelve years later, we measure 0.49$\arcsec$ and 186$^{\circ}$ for AD and 0.95$\arcsec$ and 188$^{\circ}$ for BC.

HD 152751 is also a high-order multiple system where we resolved four of its components. The primary star, HD 152751, is a spectroscopic binary, as well as it's binary companion, Wolf 629, at a separation of 72$\arcsec$. Furthermore, we resolved a companion to HD 152751 at 0.24$\arcsec$ and a wide co-moving companion to Wolf 629 at almost 300$\arcsec$ separation.

\subsection{Multiplicity Fractions}
\label{sec:multSpT}

\begin{figure}
\centering
\includegraphics[width=245pt]{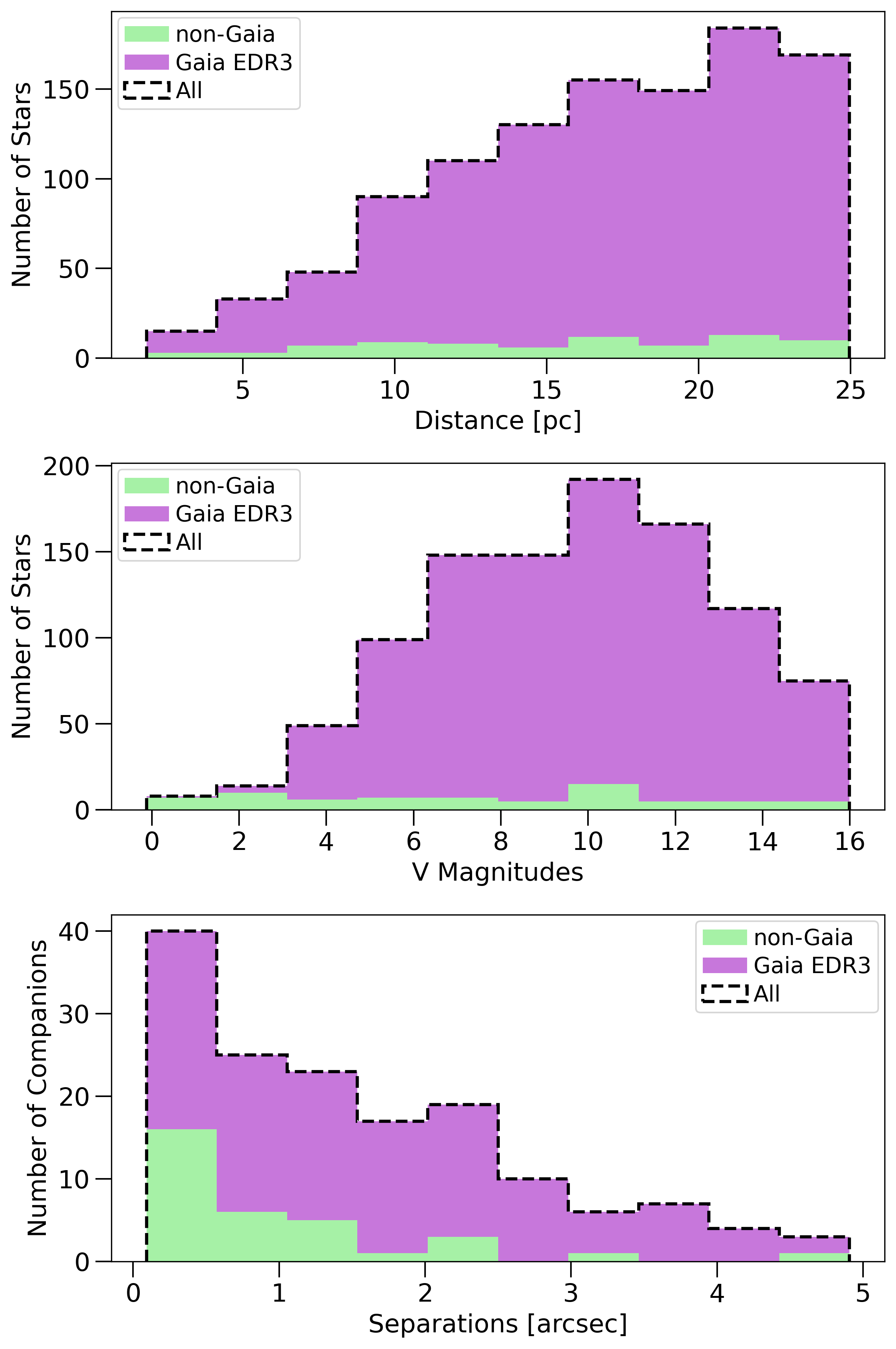}
\caption{Stacked distribution comparison between the stars found in Gaia EDR3 and those not found in Gaia where distances were found in SIMBAD. There is no clear bias in the sample distance distributions. However, there are more bright stars not found in Gaia EDR3 and possibly more bright sub-arcsecond binaries.}
\label{fig:SIMBADsample}
\end{figure}

Combining the tight companion detections from Robo-AO with the wide companion detections from Gaia EDR3 astrometry, we evaluated the multiplicity fraction of our observed sample of stars for separations ranging from a few AU (for a typical FWHM resolution of 0.15$\arcsec$ this corresponds to $\lesssim$1~AU at $<$7~pc and 3.75~AU at 25~pc) out to 10,000~AU. The multiplicity frequency was calculated only including the \NstarsGaia\ stars that were found in Gaia EDR3, where we could perform the wide companion search to complement the tight companion search with Robo-AO. Removing the stars not found in Gaia EDR3 removes a larger proportion of brighter stars. Removing the stars not in Gaia also likely removes some bright sub-arcsecond binaries, which could cause us to underestimate the multiplicity fractions. See Figure \ref{fig:SIMBADsample} for a comparison of property distributions between the Gaia and non-Gaia samples. 

\begin{figure}
\centering
\includegraphics[width=245pt]{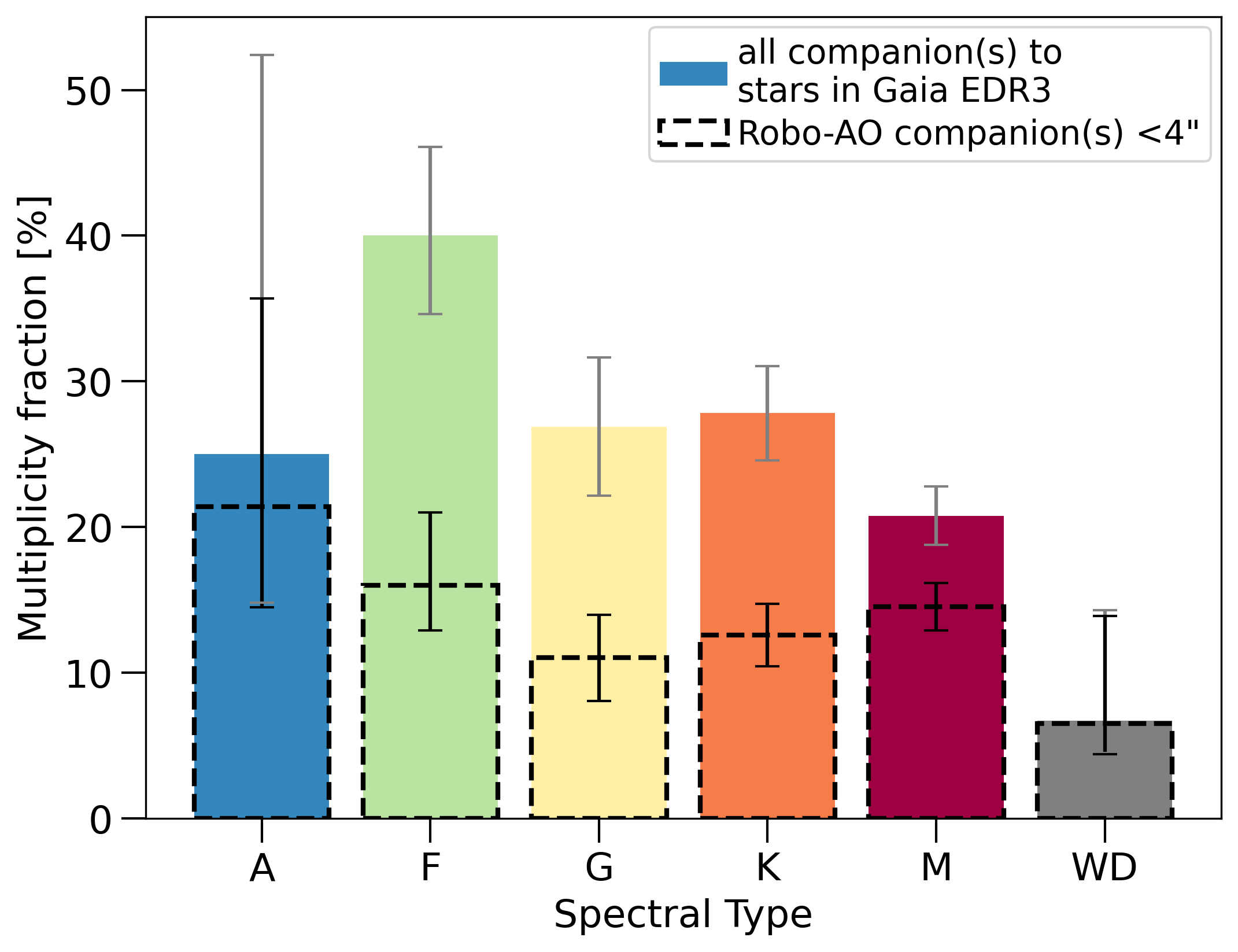}
\caption{Stellar multiplicity fractions as a function of spectral type for the sample of \NstarsGaia\ primary stars found in Gaia with both tight Robo-AO detected companions and wide co-moving companions found in Gaia (\textit{colored bars}) and for the full sample of \Nsystems\ primary stars with only tight companions detected by Robo-AO (\textit{dashed bars}).}
\label{fig:sptypefractions}
\end{figure}

Our calculated multiplicity fractions for each SpT sample are summarized in Table \ref{tab:detection_freqs} and Figure \ref{fig:sptypefractions}. We excluded systems with companions ruled out as background objects but include systems that have yet to be confirmed, as we expect the vast majority to be physically associated due to their tight separations \citep{Salama21}. The fraction of observed stars with companion(s) is \MultFracALL. We used the Poisson distribution to calculate the error bars for the samples with at least 100 stars, and numerically calculated the 1$\sigma$ uncertainties using the binomial distribution for those with less than 100 stars \citep{burgasser03}.

\begin{table*}
\centering
\caption{Companion Detection Frequencies
    \label{tab:detection_freqs}}
\begin{tabular}{c|crl|crlrl}
\hline
\textbf{SpT} & \multicolumn{3}{c}{\textbf{Full Sample}} \vline & \multicolumn{5}{c}{\textbf{Gaia Sample}} \\
\textbf{Primary} & \# Stars & \multicolumn{2}{l}{\# Stars with Comp(s) $<$4$\arcsec$} \vline & \# Stars & \multicolumn{2}{l}{\# Stars with Comp(s)} & \multicolumn{2}{l}{\# Stars with $>$1 Comp} \\
\hline
A & 14 & 3 & (21$_{-7}^{+14}$\%) & 4 & 1 & (\MultFracA) & 1 & (\MultFracHOA) \\
F & 81 & 13 & (16$_{-3}^{+5}$\%) & 70 & 28 & (\MultFracF) & 0 & $\cdots$ \\
G & 127 & 14 & (11.0\%$\pm$3.0\%) & 119 & 32 & (\MultFracG) & 3 & (\MultFracHOG) \\
K & 278 & 35 & (12.6\%$\pm$2.1\%) & 266 & 74 & (\MultFracK) & 2 & (\MultFracHOK) \\
M & 550 & 80 & (14.6\%$\pm$1.6\%) & 515 & 107 & (\MultFracM) & 13 & (\MultFracHOM) \\
WD & 31 & 2 & (7$_{-2}^{+7}$\%) & 30 & 2 & (\MultFracWD) & 0 & $\cdots$ \\
\hline
\end{tabular}
\end{table*}

The multiplicity fraction of FGK stars is estimated at 44$\pm$2\% \citep{Raghavan10} and 26.7$\pm$1.4\% \citep{Winters19} for M type stars. Our results of \MultFracFGK\ for FGK stars and \MultFracM\ for M stars are lower, however we have not incorporated spectroscopic binaries in our estimates (see Section \ref{sec:unresolved}). In addition, \cite{Winters21} found that M dwarf separations are closer, 20~AU on average, than the 50~AU average for companions to FGK stars. This makes it more challenging to resolve companions around low-mass stars. The known trend of higher multiplicity fractions for higher-mass stars \citep{Duchene13} is consistent with our results, within the error bars.

Overall, we compute a \MultFracHOAll\ higher-order multiplicity fraction. The higher order multiplicity fractions by SpT of primary stars are also shown in Table \ref{tab:detection_freqs} for the full sample with only tight ($<$4$\arcsec$) companions, and for the Gaia EDR3 sample with all its companions. \cite{Duchene13} estimate high order multiplicity fractions of $\approx$11\% for solar-type stars (FGK types), whereas we calculate \MultFracHOFGK. However, they also include spectroscopic binaries. \cite{Winters19} estimate a high order multiplicity rate of $\approx$5\% for M dwarfs, which is slightly higher than our estimate of \MultFracHOM. The \cite{Winters19} study was also conducted on a volume-limited sample out to 25~pc and searched for companions from 2$\arcsec$--300$\arcsec$, however they also included known sub-arcsecond companions from the literature and because all of their targets within 10~pc had been observed with high-resolution techniques, they applied the same $<2\arcsec$ multiplicity rate to stars with distances 10--25~pc.

\subsubsection{Correcting for Unresolved Companions}
\label{sec:unresolved}

We cross-matched our Gaia sample of stars with the ninth catalog of spectroscopic binary orbits (SB9; \citealt{Pourbaix04}\footnote{version:2021-03-02 10:50:23.278733383 +0100}) \new{and a catalog of eclipsing variables from \cite{Malkov06}} in order to correct our multiplicity fractions for unresolved binaries. We found that \NSBs\ of our stars are spectroscopic binaries \new{and an additional \NEBs\ are eclipsing binaries}, which adds \new{\NSBsnewBinaries}\ new binaries and \new{\NSBsnewHOs}\ higher-order multiple systems to our sample. After applying this correction, our FGK and M multiplicity fractions increase to \MultFracFGKSBs\ and \MultFracMSBs, respectively, and to \new{\MultFracHOFGKSBs}\ and \MultFracHOMSBs\ for their corresponding higher-order multiplicity fractions.

Stars with high RUWE values ($>$2.0) but with no detected companion within 2$\arcsec$ and/or with significant accelerations ($\chi^2 >$11.8) but with no detected companion within 300~AU, likely have unresolved tight companions. These separation cutoffs were chosen based on Figures \ref{fig:comp_seps_RUWEs} \& \ref{fig:acc_chisqs_seps}, and with the assumption that a companion's separation on-sky in arcseconds is what impacts Gaia's ability to resolve the pair and a companion's true separation in AU is what affects the acceleration of its primary star. We thus created a sample of ``potentially unresolved" multiple systems. In addition, following \cite{Vrijmoet20}, who determined that 75\% of stars with RUWEs$>$2.0 have unresolved companions, we removed 25\% of the counts of our ``potentially unresolved" sample to not over-correct our multiplicity fraction calculations. With these assumptions, we would expect an additional \NcompsUNRESOLVED\ stars to have unresolved tight companions, and an additional \NhigherOrdersUNRESOLVED\ higher-order systems where we have detected the wider companion only. If this were the case, our multiplicity fractions become \MultFracFGKunrs\ for FGK stars and \MultFracMunrs\ for M type stars and \MultFracHOFGKunrs\ and \MultFracHOMunrs\ for their corresponding higher-order fractions. This method likely over-corrects for low-mass stars, which are fainter and thus are more likely to have issues with the astrometry fits based on their faintness without necessarily being caused by the presence of a companion. Another source of error is that the estimate from \cite{Vrijmoet20} of 75\% of stars with RUWEs$>$2.0 having unresolved companions is based on Gaia DR2 and would likely increase with improved astrometry from Gaia EDR3. 

Both of these correction methods yield multiplicity fractions that are closer than our raw multiplicity fractions, to the current estimates from the literature. Another source of error, particularly affecting the fainter stars in our sample, is a bias towards detecting binaries due to an increase in brightness of the system. Faint stars that have been more challenging to detect and to measure their parallaxes, especially prior to Gaia, will appear brighter if there are two stars and thus have been more likely to be counted in nearby stellar censuses.

\section{Conclusion}
\label{sec:conclusion}

Our objective with the Robo-AO Solar Neighborhood Survey is to provide a uniform sample of the nearby stellar population, allowing more accurate multiplicity statistics across spectral types and stellar populations. We observed \Nsystems\ stellar systems within 25~pc across all spectral types. Our main findings are:

\begin{itemize}
    \item We detected \Ncomps\ companions, \NcompsRAO\ are $<$4$\arcsec$ companions detected with high-resolution imaging using the Robo-AO adaptive optics instrument, and \NcompsGaiaDet\ were detected using Gaia EDR3 astrometry, spanning out to 10,000 AU. \NcompsNew\ are new companion detections.
    \item We detected \Nbinaries\ binary systems, \Ntriples\ triple systems, and \Nquads\ higher-order systems.
    \item We evaluated the Gaia EDR3 companion recovery and correlation with RUWE values and found that most companions within 1$\arcsec$ were not detected by Gaia but the host stars have a large ($>$1.4) RUWE value.
    \item We found that \FracAccTightCompsSig\ of the stars with companions within 4$\arcsec$ had significant acceleration measurements in the HGCA catalog, compared to \FracAccWideCompsSig\ for stars with wide ($>$4$\arcsec$) companions, and \FracAccNOCompsSig\ for stars with no detected companions. 
    \item From these results, we have calculated the raw multiplicity fractions of stars within 25~pc to be \MultFracFGK\ for FGK stars and \MultFracM\ for M stars, which we estimate to increase to \MultFracFGKunrs\ and \MultFracMunrs\ for FGK and M type stars if we account for unresolved companions. 
\end{itemize}

Surveys such as ours with Robo-AO allow us to probe for tight companions not resolvable with non-AO observations. This provides a more complete census of the number of stellar multiple systems among our nearest neighbors and will allow for more detailed statistics on the characteristic distribution of these stellar multiple systems. Follow-up studies constraining the orbits of these companions will provide important information on the dynamics of these systems and mass calculations of each stellar component in the systems. These orbital dynamics, mass measurements, and more detailed statistics of multiplicity fractions will help constrain stellar evolution and formation models, further informing us on our understanding of the stellar populations in our galaxy.

\acknowledgments

The Robo-AO system is supported by collaborating partner institutions, the California Institute of Technology and the Inter-University Centre for Astronomy and Astrophysics, and by the National Science Foundation under Grant Nos. AST-0906060, AST-0960343, and AST-1207891, by the Mount Cuba Astronomical Foundation, and by a gift from Samuel Oschin. 

We are grateful to the Palomar Observatory staff for their ongoing support of Robo-AO on the 60 inch telescope, particularly S. Kunsman, M. Doyle, J. Henning, R. Walters, G. Van Idsinga, B. Baker, K. Dunscombe and D. Roderick.

This research was funded in part by grants to M. Liu from the Gordon and Betty Moore Foundation (Grant GBMF8550) and the National Science Foundation (AST-1518339)

JGW is supported by the National Aeronautics and Space Administration under Grant No.~80NSSC18K0476 issued through the XRP Program.

This research has made use of the SIMBAD database, operated by Centre des Donn\'ees Stellaires (Strasbourg, France), and bibliographic references from the Astrophysics Data System maintained by SAO/NASA.

This work has made use of data from the European Space Agency (ESA) mission {\it Gaia} (\url{https://www.cosmos.esa.int/gaia}), processed by the {\it Gaia} Data Processing and Analysis Consortium (DPAC, \url{https://www.cosmos.esa.int/web/gaia/dpac/consortium}). Funding for the DPAC has been provided by national institutions, in particular the institutions participating in the {\it Gaia} Multilateral Agreement.

This research has made use of the Washington Double Star Catalog maintained at the U.S. Naval Observatory.

%

\vspace{5mm}
\facilities{PO:1.5m (Robo-AO)}






\bibliography{sample63}{}
\bibliographystyle{aasjournal}

\clearpage
\begin{longrotatetable}
\begin{deluxetable*}{lllcccccccccc}
\tablecaption{All Observed Targets}
\tablewidth{700pt}
\tabletypesize{\scriptsize}
\tablehead{
\colhead{Target ID} & \colhead{Target ID} & \colhead{Source ID} & \colhead{R. A.} & \colhead{Dec.} & \colhead{\textit{V}} & \colhead{\textit{G}} & \colhead{Distance} & \colhead{SpT} & \colhead{Obs. Date} & \colhead{\textit{Gaia}} & \colhead{HIP-\textit{Gaia}} & \colhead{Comp.}\\ 
\colhead{Robo-AO} & \colhead{SIMBAD} & \colhead{Gaia EDR3} & \colhead{J2000} & \colhead{J2000} & \colhead{(mag)} & \colhead{(mag)} & \colhead{(pc)} & & \colhead{(UTC)} & \colhead{RUWE} & \colhead{prim. $\chi^2_{acc}$} & \colhead{Candidate(s)?} }
\startdata
0005+4548 A & HD     38A & 386653851004022144 & 00:05:41.02 & +45:48:43.55 & 8.23 & 8.25 & 11.52 $\pm$ 0.00 [G] & K6V [S] & 2012 Jul 17 & 1.2 & 6209.3 & yes\\  
0006+5826 A & HD    123A & 423075173680043904 & 00:06:15.81 & +58:26:12.11 & 6.11 & 6.22 & 20.79 $\pm$ 0.02 [G] & G3V [S] & 2012 Jul 17 & 1.8 & 45358.3 & yes\\  
0006+2901 A & HD    166 & 2860924621205256704 & 00:06:36.78 & +29:01:17.41 & 6.21 & 5.88 & 13.77 $\pm$ 0.01 [G] & G8 [S] & 2012 Jul 17 & 0.9 & 3.1 & $\cdots$\\  
0006-0732 A & G 158-27 & 2441630500517079808 & 00:06:43.20 & -07:32:17.02 & 13.77 & 11.77 & 4.85 $\pm$ 0.00 [G] & M5.5V [S] & 2013 Jul 27 & 1.3 & $\cdots$ & $\cdots$\\  
0007+2914 A & G 130-43 & 2860957125517813248 & 00:07:26.71 & +29:14:32.77 & 14.15 & 12.95 & 21.11 $\pm$ 0.01 [G] & M3.5V [S] & 2012 Jul 17 & 1.1 & $\cdots$ & $\cdots$\\  
0008+1725 A & GJ  3008 & 2772804845911842944 & 00:08:27.28 & +17:25:27.46 & 10.8 & 10.03 & 21.74 $\pm$ 0.01 [G] & K7V [S] & 2012 Oct 09 & 1.0 & 2.7 & $\cdots$\\  
0009+5908 A & bet Cas & $\cdots$ & 00:09:10.6851 & +59:08:59.212 & 2.36 & $\cdots$ & 16.78 [S] & F2III [S] & 2012 Jul 17 & $\cdots$ & $\cdots$ & $\cdots$\\  
0011+5908 A & LSPM J0011+5908 & 423027104407576576 & 00:11:31.82 & +59:08:39.88 & 15.68 & 13.47 & 9.31 $\pm$ 0.00 [G] & M6.0V [S] & 2012 Aug 05 & 1.1 & $\cdots$ & $\cdots$\\
\enddata
\tablenotetext{}{[G] from Gaia EDR3; [S] from SIMBAD}
\label{tab:obs_all}
\end{deluxetable*}
\end{longrotatetable}

\clearpage
\begin{longrotatetable}
\startlongtable
\begin{deluxetable*}{llcccccccccccc}
\tablecaption{Companion Candidate Measurements}
\tablewidth{700pt}
\tabletypesize{\scriptsize}
\tablehead{
\colhead{Target ID} & \colhead{Source ID} & \colhead{Detection} & \colhead{Epoch} & \colhead{Separation} & \colhead{Separation} & \colhead{P. A.} & \colhead{Filter} & \colhead{Contrast} & \colhead{Detection} & & & \colhead{Phys.}  & \\ 
\colhead{Robo-AO} & \colhead{Gaia EDR3} & \colhead{Source} & \colhead{(Julian date)} & \colhead{($\arcsec$)} & \colhead{(AU)} & \colhead{($^{\circ}$)} & & \colhead{(mag)} & \colhead{S/N} & \colhead{$\Delta \pi/\pi$} & \colhead{$\Delta \mu/\mu$} & \colhead{assoc.} & \colhead{$\mathcal{R}$}}
\startdata
0005+4548 AB $^{WDS}$ & 386653747925624576 & Gaia EDR3 & $\cdots$ & 6.03 & 69.50 $\pm$ 0.02 & 188.4 & \textit{G} & 0.05 & $\cdots$ & 10$^{-4}$ & 10$^{-2}$ & yes & 10$^{-8}$\\  
0005+4548 AC $^{Int4}$ & 386655019234959872 & Gaia EDR3 & $\cdots$ & 328.48 & 3778.66 $\pm$ 0.74 & 253.8 & \textit{G} & 0.83 & $\cdots$ & 10$^{-3}$ & 10$^{-2}$ & yes & $\cdots$\\  
0006+5826 AB $^{R10}$ & 423075173674854528 & Robo-AO & 2012.5420580 & 1.51 $\pm$ 0.06 & 32.11 $\pm$ 1.28 & 352.0 $\pm$ 2.3 & \textit{i'} & 0.77 $\pm$ 0.11 & 217 & 10$^{-2}$ & 0.2 & yes & 10$^{-6}$\\  
0013+6919 AB $^{W19}$ & $\cdots$ & Robo-AO & 2012.5447009 & 0.83 $\pm$ 0.06 & 17.04 $\pm$ 1.23 & 89.4 $\pm$ 4.1 & \textit{i'} & 0.33 $\pm$ 0.13 & 110 & $\cdots$ & $\cdots$ & $\cdots$ & $\cdots$\\  
0013+8039 AB $^{W19}$ & 565169459376414592 & Gaia EDR3 & $\cdots$ & 12.76 & 245.40 $\pm$ 0.28 & 126.2 & \textit{G} & 4.20 & $\cdots$ & 10$^{-5}$ & 10$^{-2}$ & yes & 10$^{-6}$\\  
0016+1951 AB $^{W19}$ & 2797745549200773120 & Gaia EDR3 & $\cdots$ & 25.14 & 386.15 $\pm$ 0.23 & 58.5 & \textit{G} & 0.98 & $\cdots$ & 10$^{-4}$ & 10$^{-2}$ & yes & 10$^{-11}$\\  
0018+4401 AB $^{W19}$ & 385334196532776576 & Gaia EDR3 & $\cdots$ & 34.35 & 122.36 $\pm$ 0.01 & 65.5 & \textit{G} & 2.47 & $\cdots$ & 10$^{-5}$ & 10$^{-2}$ & yes & 10$^{-6}$\\  
0023+2418 AB $^{WDS}$ & 2801216123294143488 & Robo-AO & 2012.5419523 & 2.26 $\pm$ 0.06 & 4636.00 $\pm$ 797.87 & 254.3 $\pm$ 1.5 & \textit{i'} & 3.24 $\pm$ 0.09 & 160 & 1.0 & 1.0 & no & $\cdots$\\
\enddata
\tablenotetext{}{$^{W19}$ Found in \cite{Winters19}.}
\tablenotetext{}{$^{R10}$ Found in \cite{Raghavan10}.}
\tablenotetext{}{$^{WDS}$ Found in the Washington Double Star catalog.}
\tablenotetext{}{$^{Int4}$ Found in the Fourth Catalog of Interferometric Measurements of Binary Stars.}
\label{tab:comps_measurements}
\end{deluxetable*}
\end{longrotatetable}

\clearpage
\startlongtable
\begin{deluxetable*}{llccccccc|llccccccc}
\tablecaption{5$\sigma$ Detection Limits}
\tablewidth{700pt}
\tabletypesize{\scriptsize}
\tablehead{
\textbf{Target ID} & \colhead{Obs} & \multicolumn{7}{c}{Limit (mag) at Separation}  & \textbf{Target ID} & \colhead{Obs} & \multicolumn{7}{c}{Limit (mag) at Separation} \\ 
\textbf{Robo-AO} & \colhead{Quality} & \colhead{0$\farcs$5} & \colhead{1$\farcs$0} & \colhead{1$\farcs$5} & \colhead{2$\farcs$0} & \colhead{2$\farcs$5} & \colhead{3$\farcs$0} & \colhead{3$\farcs$5} & \textbf{Robo25} & \colhead{Quality} & \colhead{0$\farcs$5} & \colhead{1$\farcs$0} & \colhead{1$\farcs$5} & \colhead{2$\farcs$0} & \colhead{2$\farcs$5} & \colhead{3$\farcs$0} & \colhead{3$\farcs$5} }
\startdata
0005+4548 A & high & 1.52 & 3.19 & 4.51 & 5.36 & 6.0 & 6.36 & 6.63 & 1147+5058 A & high & 1.29 & 2.69 & 3.92 & 4.88 & 5.57 & 6.07 & 6.41\\ 
0006+2901 A & high & 0.93 & 2.45 & 3.67 & 4.66 & 5.41 & 6.01 & 6.37 & 1147+0048 A & high & 1.53 & 2.93 & 4.12 & 5.09 & 5.75 & 6.29 & 6.67\\ 
0006-0732 A & high & 2.47 & 4.22 & 5.31 & 5.86 & 6.11 & 6.23 & 6.3 & 1149+1434 A & high & 0.0 & 0.3 & 1.16 & 1.93 & 2.62 & 3.2 & 3.7\\ 
0008+1725 A & high & 1.23 & 3.06 & 4.29 & 5.14 & 5.77 & 6.26 & 6.48 & 1150+0145 A & high & 1.04 & 2.4 & 3.63 & 4.57 & 5.3 & 5.92 & 6.35\\ 
0009+5908 A & high & 0.73 & 2.2 & 3.26 & 4.08 & 4.78 & 5.49 & 6.06 & 1150+4822 A & high & 1.71 & 3.12 & 4.26 & 4.97 & 5.36 & 5.54 & 5.69\\ 
0012+5025 A & low & 1.41 & 2.52 & 2.91 & 3.09 & 3.2 & 3.21 & 3.24 & 1151+3516 A & high & 0.15 & 1.49 & 2.66 & 3.58 & 4.3 & 4.93 & 5.32\\ 
0013+8039 A & high & 1.11 & 2.97 & 4.34 & 5.32 & 5.91 & 6.33 & 6.64 & 1153-0722 A & high & 0.1 & 1.14 & 2.24 & 3.12 & 3.76 & 4.28 & 4.61\\ 
0016+1951 A & high & 0.46 & 1.87 & 3.07 & 3.94 & 4.55 & 4.95 & 5.19 & 1154+0948 A & high & 0.37 & 1.79 & 2.92 & 3.71 & 4.19 & 4.49 & 4.74\\
\enddata
\label{tab:roboaolist}
\end{deluxetable*}

\clearpage
\begin{longrotatetable}
\startlongtable
\begin{deluxetable*}{llccccccccccccc}
\tablecaption{Companion Candidate Measurements for Stars $>$25 pc}
\tablewidth{700pt}
\tabletypesize{\scriptsize}
\tablehead{
\colhead{Prim. Source ID} & \colhead{Comp. Source ID} & \colhead{Distance} & \colhead{Detection} & \colhead{Epoch} & \colhead{Separation} & \colhead{Separation} & \colhead{P. A.} & \colhead{Filter} & \colhead{Contrast} & \colhead{Detection} & & & \colhead{Phys.}  & \\ 
\colhead{Gaia EDR3} & \colhead{Gaia EDR3} & \colhead{(pc)} & \colhead{Source} & \colhead{(Julian date)} & \colhead{($\arcsec$)} & \colhead{(AU)} & \colhead{($^{\circ}$)} & & \colhead{(mag)} & \colhead{S/N} & \colhead{$\Delta \pi/\pi$} & \colhead{$\Delta \mu/\mu$} & \colhead{assoc.} & \colhead{$\mathcal{R}$}}
\startdata
435570539206096640 & $\cdots$ & 25.05 $\pm$ 0.05 [G] & Robo-AO & 2012.5831972 & 0.73 $\pm$ 0.06 & 18.32 $\pm$ 1.50 & 193.9 $\pm$ 4.7 & \textit{i'} & 2.43 $\pm$ 0.14 & 72 & $\cdots$ & $\cdots$ & $\cdots$ & $\cdots$\\  
9178383766474496 & $\cdots$ & 25.14 $\pm$ 0.02 [G] & Robo-AO & 2012.6679375 & 0.23 $\pm$ 0.08 & 5.84 $\pm$ 2.07 & 81.5 $\pm$ 17.4 & \textit{i'} & 2.28 $\pm$ 0.16 & 14 & $\cdots$ & $\cdots$ & $\cdots$ & $\cdots$\\  
9178383766474496 & 9178383766474624 & 25.14 $\pm$ 0.02 [G] & Robo-AO & 2012.6679375 & 2.09 $\pm$ 0.12 & 52.58 $\pm$ 2.92 & 213.4 $\pm$ 2.9 & \textit{i'} & 0.16 $\pm$ 0.11 & 246 & $\cdots$ & $\cdots$ & $\cdots$ & $\cdots$\\  
3199115603133450880 & 3199115603133451136 & 36.09 $\pm$ 0.03 [G] & Robo-AO & 2012.7009272 & 2.95 $\pm$ 0.06 & 106.32 $\pm$ 2.17 & 282.1 $\pm$ 1.2 & \textit{i'} & 0.82 $\pm$ 0.21 & 269 & 10$^{-3}$ & 10$^{-3}$ & yes & 10$^{-37}$\\  
3413754360367271168 & 3413754360367270272 & 31.78 $\pm$ 0.08 [G] & Gaia EDR3 & $\cdots$ & 9.45 & 300.27 $\pm$ 0.76 & 19.4 & \textit{G} & 7.36 & $\cdots$ & 10$^{-3}$ & 10$^{-2}$ & yes & 10$^{-6}$\\  
3379666746830064384 & 3379666751124980352 & 1383.13 $\pm$ 255.77 [G] & Robo-AO & 2012.7720618 & 2.49 $\pm$ 0.06 & 3442.82 $\pm$ 642.05 & 155.9 $\pm$ 1.4 & \textit{i'} & 5.40 $\pm$ 0.58 & 90 & 1.0 & 1.0 & no & $\cdots$\\  
950395658954590080 & 950395654664758272 & 25.05 $\pm$ 0.01 [G] & Gaia EDR3 & $\cdots$ & 37.90 & 949.16 $\pm$ 0.49 & 6.2 & \textit{G} & 1.61 & $\cdots$ & 10$^{-3}$ & 10$^{-2}$ & yes & 10$^{-6}$\\  
870758684590736896 & 870758684589320448 & 27.17 $\pm$ 0.21 [G] & Robo-AO & 2013.0561190 & 2.40 $\pm$ 0.06 & 65.32 $\pm$ 1.71 & 138.2 $\pm$ 1.4 & \textit{i'} & 4.22 $\pm$ 0.26 & 304 & 10$^{-2}$ & 10$^{-2}$ & yes & 10$^{-8}$\\
\enddata
\label{tab:beyond25pc}
\end{deluxetable*}
\end{longrotatetable}

\end{document}